\documentclass[a4paper,fleqn]{cas-dc}
\pdfoutput=1
\usepackage[authoryear,longnamesfirst]{natbib}
\usepackage{graphicx}

\begin{document}
\let\WriteBookmarks\relax
\def\floatpagepagefraction{1}
\def\textpagefraction{.001}

\shorttitle{Hierarchical Clustering in Astronomy}

\shortauthors{Yu \& Hou}

\title [mode = title]{Hierarchical Clustering in Astronomy}                      

\author[1]{Heng Yu}

\cormark[1]

\address[1]{Department of Astronomy, Beijing Normal University, 100875, Beijing,China}
% \affiliation[1]{organization={Department of Astronomy, Beijing Normal University},
%     %addressline={Radarweg 29},
%     city={Beijing},
%     % citysep={}, % Uncomment if no comma needed between city and postcode
%     postcode={100875},
%     % state={},
%     country={China}}

\author[1]{Xiaolan Hou}[style=chinese]

\begin{abstract}
Hierarchical clustering is a common algorithm in data analysis. It is unique among many clustering algorithms in that it draws dendrograms based on the distance of data under a certain metric, and group them. It is widely used in all areas of astronomical research, covering various scales from asteroids and molecular clouds, to galaxies and galaxy cluster. This paper systematically reviews the history and current status of the development of hierarchical clustering methods in various branches of astronomy. These applications can be grouped into two broad categories, one revealing the intrinsic hierarchical structure of celestial systems and the other classifying large samples of celestial objects automatically. By reviewing these applications, we can clarify the conditions and limitations of the hierarchical clustering algorithm, and make more reasonable and reliable astronomical discoveries.

\end{abstract}


\maketitle

\section{Introduction}

The classification of data according to their different properties is a basic strategy of scientific research. With the progress of research techniques, these classification methods are gradually abstracted to form a specialized field -- cluster analysis \citep{Everitt2011}. 
Most of the classification and grouping work in astronomy can find corresponding algorithms in this field \citep{1996Babu, Ivezic2014}. 
Emerging techniques in this field are also often borrowed and developed by astronomers.

Hierarchical clustering, or hierarchical cluster analysis (HAC) is such a data analysis algorithm. It is to quantify the distance (or similarity) between data according to some predefined metric, and use dendrogram to present these distances and reveal hierarchical relationships. In 1948, Danish botanist \cite{Sorensen1948} proposed the original form of this algorithm. After years of exploration and improvement by many researchers, a mature hierarchical clustering algorithm has been developed \citep{Lance1967, Gordon1987}. It does not require users to restrict a priori properties, such as the number of clusters and distribution functions of the sample in advance. So it is an unsupervised learning algorithm in the field of machine learning. The unique advantage of hierarchical clustering is that it does not only provide the final grouping results like other algorithms, but also organizes all the data in a dendrogram that can preserve the complete hierarchy and clustering relationships between data. It gives users the opportunity to select appropriate groups of features from the dendrogram based on different thresholds, and it preserves the associations between the groups.

The universe is hierarchical, because similar organized structures can be seen at different cosmic scales, as stars converge into clusters and galaxies, galaxies form galaxy clusters, and galaxy clusters are interconnected to form the large-scale structure of the universe. Although their spatial scales are very different, their formation and evolution are all dominated by gravity, and they have a similar hierarchical structure, which is suitable for graphical presentation. Hierarchical clustering caught the attention of astronomers as early as 1978, French astronomer Materne introduced it into the field of astronomy, to figure galaxy groups in nearby galaxies \citep{1978A&A....63..401M}. Later, this algorithm has been widely used in all branches of astronomy.

This paper systematically introduces the history and current status of the development of hierarchical clustering algorithms in different application scenarios in the field of astronomy. In Section \ref{sec:method}, we describe the basic concepts and steps of algorithms; in Section \ref{sec:appl}, we specifically describe the application of the algorithm in different research branches; Finally, we summarize and discuss its current status and development prospects.

\section{Algorithm}
\label{sec:method}

The hierarchical cluster analysis have two approaches: one is the "bottom-up" agglomerative method, and the other is the "up-bottom" divisive method. The former starts from a single point of data and merges adjacent points step by step according to a given rule until all data points are combined into one class. While the latter is to treat the whole data set as a whole first and partition it according to certain rules until all data points are separated from each other. These two pathways are mutually inverse operations, and the dendrogram obtained under the same rules are the same. Because that 
agglomerative hierarchical clustering is more commonly used, it is used as an example below.

\subsection{Steps}

The main steps of agglomerative hierarchical clustering are as follows:

\begin{itemize}
 \item[(a)] First, each data point is treated as a set, and the distance between sets is calculated according to a specific metric;
 
 \item[(b)] Find the two sets with the smallest distance according to a specific linkage, merge them into a new set, and recalculate the distance between sets. Repeat this step until only one set remains;
 
 \item[(c)] Draw a dendrogram according to the order in which sets are merged;
 
 \item[(d)] Select a suitable trim threshold to obtain the cluster classification from the dendrogram.
\end{itemize}

During the process, the most critical design is how to define metric and linkage. Common examples usually adopt the simplest Euclidean distance as the metric. But in practical astronomical applications, the data is multidimensional and has different units and scales. It means that simple superposition is not possible. One way is to normalize them to dimensionless data. However, the normalization parameters are equivalent to the weight of data. They need to be chosen carefully according to experience. Even if feasible, they still face the dilemma of lacking a reasonable physical explanation. Therefore, in practical applications, it is crucial to design an  appropriate metric based on the specific physical problem and available data.

\begin{figure*}[htp]
\centering
\includegraphics[width=0.95\textwidth]{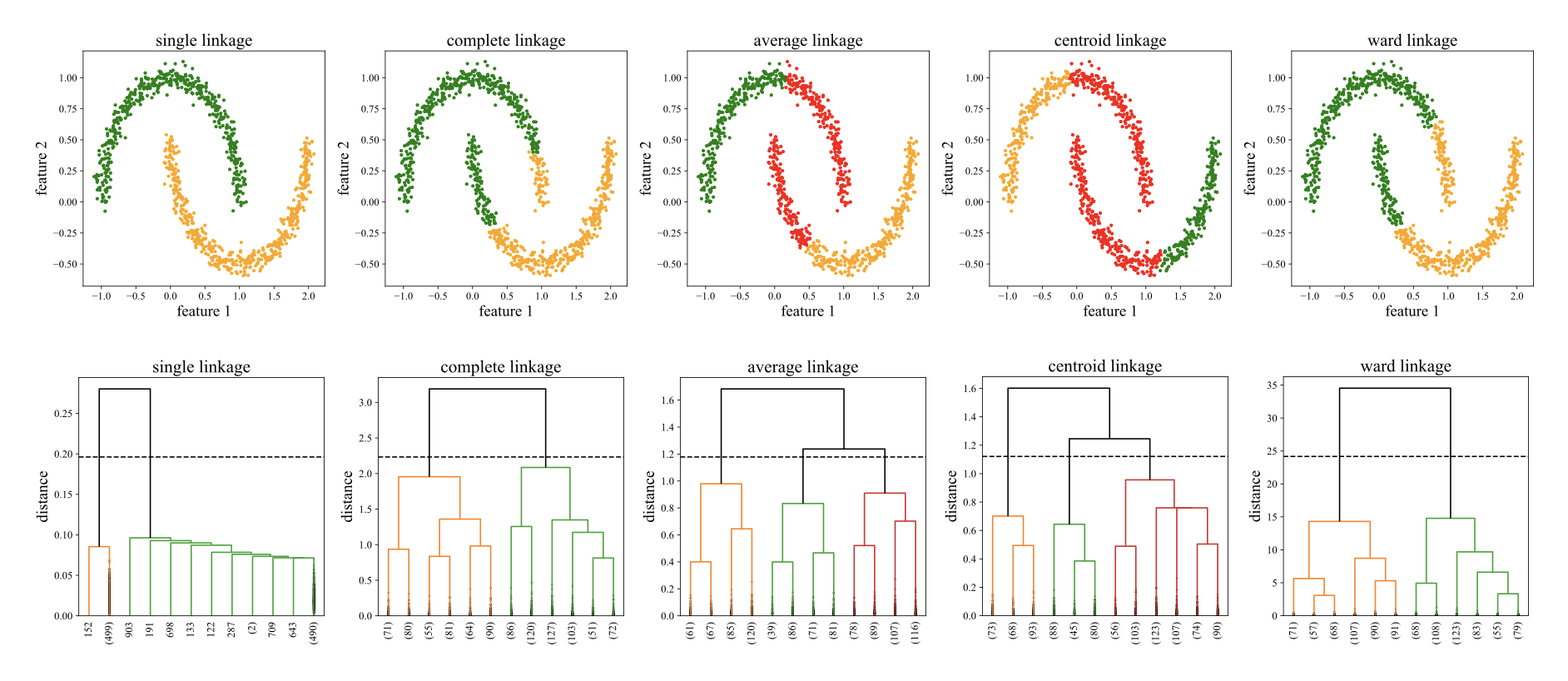}
\caption{Comparison of the results of hierarchical clustering with different connections. This figure is adapted from \cite{2019Baron} with the Python module scipy.cluster.hierarchy.}
\label{fig:linkage}
\end{figure*}

Second, there are various ways to link sets\citep{Everitt2011}, usually includes:
\begin{itemize}
\item Single linkage. The sets to be linked are determined based on the size of the distance between the nearest points of the sets. Because considering only the nearest group of members between sets will merge the whole set preferentially due to the close distance between individual members, the regular shape and distribution of the group cannot be guaranteed. This is known as the chaining effect and is advantageous in detecting irregular structures\citep{2019Baron}.
\item Complete linkage. Determine the set to be joined based on the size of the distance between the farthest points of the sets. By considering only the farthest set of members between sets, the merging hierarchy of the entire set can be severely affected by individual outliers.
\item Average linkage. Determines the set to be connected based on the average size of the distances between all points in the set; neutralizes the effect of single-point and full-set connections, tending to produce a uniformly distributed set.
\item Median linkage. Determine the set to be connected based on the median size of the distance between all points in the set; close to the effect of the mean connection and less susceptible to individual outliers.
\item Centroid linkage. Determine the set to be connected based on the distance between the centroids of the sets.
\item Ward linkage. Determine the set to be connected based on the change in variance between the original set and the new set after the connection \citep{1978A&A....63..401M}; which can better separate two noisy data sets.
\end{itemize}
Even with the same metric, different connection methods can lead to very different results, as shown in the Fig.\ref{fig:linkage}, which just shows a special case, and should not be taken as a general evaluation of these connections. In fact, these connections are suitable for different scenarios. The selection of the specific connection method needs to be considered in conjunction with the study object characteristics and metric design.

Among these linkage methods, both the single linkage and the complete linkage only need to calculate the distance between two data for one time, the time complexity and space complexity of these two algorithms are $O(N^{2})$ \citep{Sibson1973SLINKAO,Defays77}. But for the average-linkage, median-linkage, centroid-linkage and  Ward-linkage, because the distance matrix needs to be updated and stored at each iteration, the time complexity  are $O(N^{3})$, and the space complexity are $O(N^{2})$.

With the dendrogram, in principle, the cluster classification can be obtained by choosing a reasonable threshold to trim the tree.
Therefore, most of the descriptions of hierarchical clustering in the literature stop at the step of drawing a dendrogram and do not involve the operation of extracting clusters from the tree diagram.
In practical applications, clusters are not always clearly visible in the dendrogram. Choosing a reasonable trim threshold is crucial.
Manual threshold selection is inevitably influenced by subjective factors such as the user's empirical preferences. Thus a clear trimming rule is directly related to the stability of results and the automation of the method. However, there is no universal method for trimming dendrograms. They need to be designed according to specific requirements.

\subsection{Dendrogram}

The signature result of the hierarchical clustering algorithm is the dendrogram. Each node of it is derived from a single set merge operation, include two member branches, thus it is called binary tree in computer science. The node that contains all members is called the root; while the node representing a single original member at the other end is called the leaf. The x-axis position of these leaves in the dendrogram only represent their relative linking sequence. There 
are multiple topologically equivalent ways of drawing the same dendrogram, which represent exactly the same group relationships. The y-axis of the dendrogram usually uses the generation order of nodes or the metric distance, and it can actually be set flexibly according to actual needs.
\begin{figure}[htp]
\centering
\includegraphics[width=0.48\textwidth]{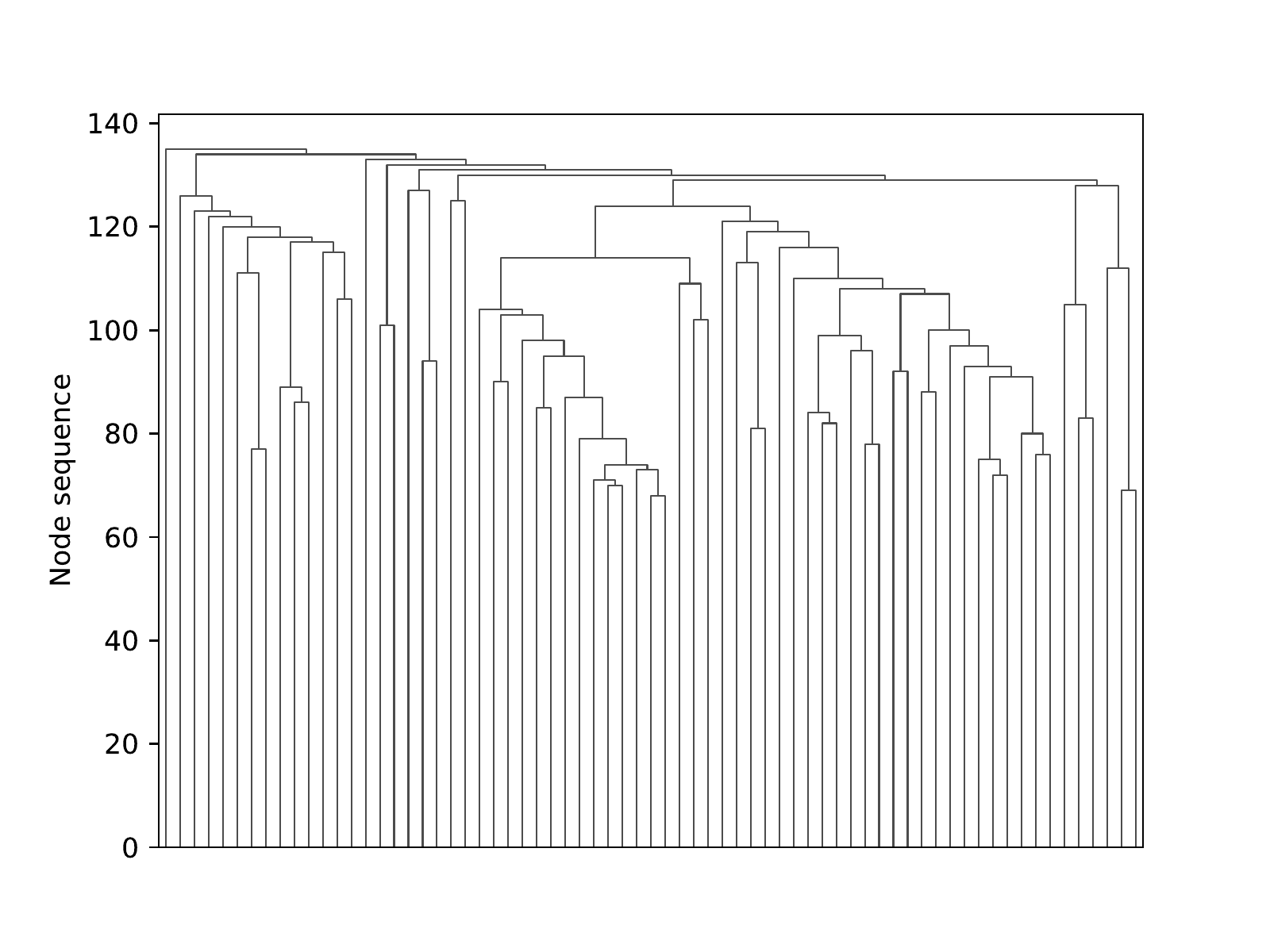}
\includegraphics[width=0.48\textwidth]{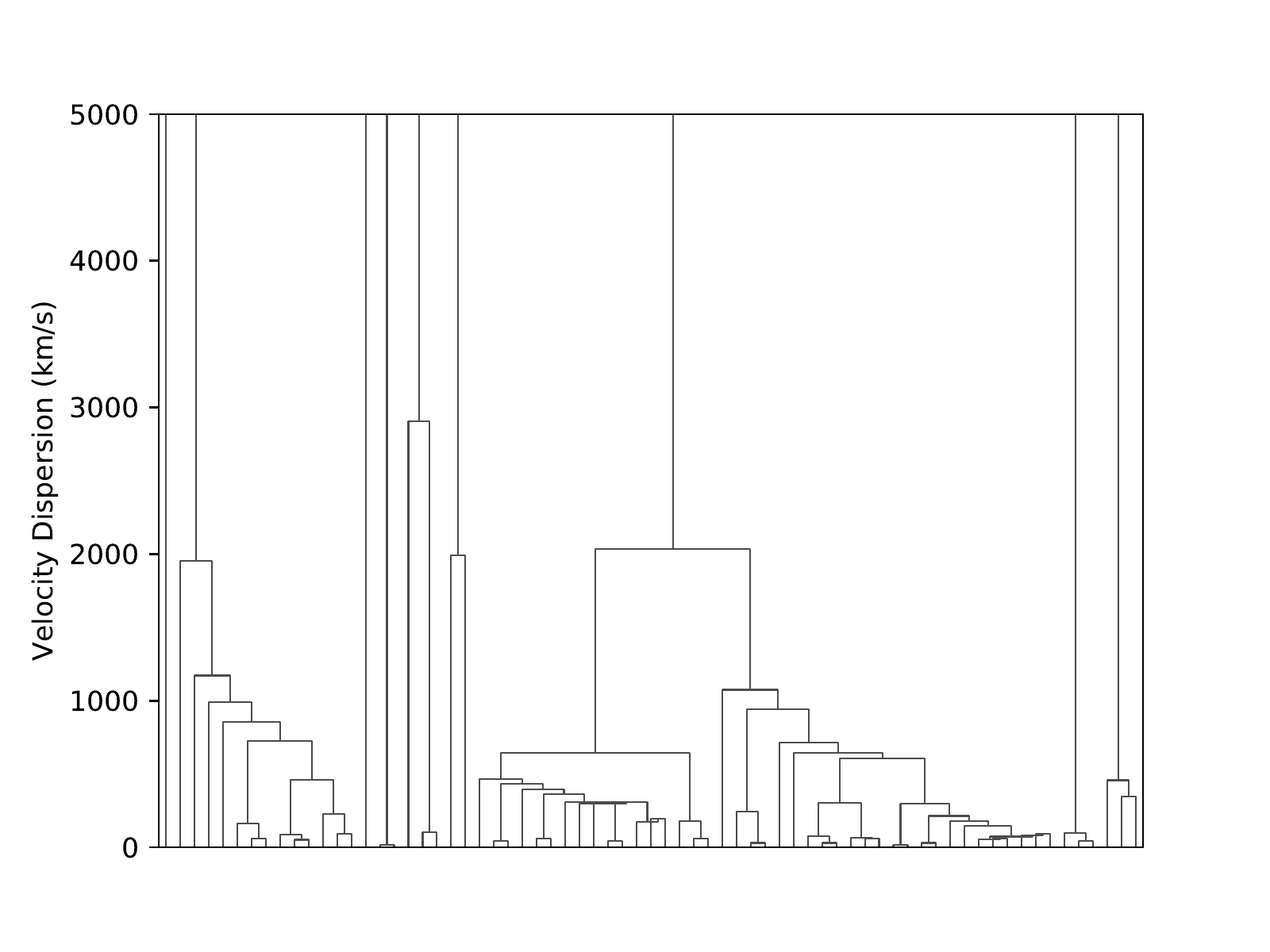}

\caption{Dendrogram of optical galaxies in the field of view of galaxy cluster MACS0358 with the data of \citet{MACS0358}. The upper panel is plotted with the node ordering as the y-axis, and the lower panel is plotted with the velocity dispersion of each node member as the y-axis. These figures are plotted with the Python module scikit-learn.}
\label{fig:dendrogram}
\end{figure}

Using different properties as the y axis will make the dendrogram look different. Fig.\ref{fig:dendrogram} is the dendrogram of optical galaxies in the field view of galaxy cluster MACS0358 \citep{wang2021}, plotted with  the projection binding between galaxies as the metric(see Section \ref{label:gcluster}). The upper panel adopts the nodal merger order as the y-axis. It shows the tightness of the gravitational connection between galaxies as it is,  but it is difficult to find the trim threshold from it due to the lack of physical meaning. Considering gravitationally-bound galaxy cluster members will exhibit relatively stable values of velocity dispersion, we can adopt the velocity dispersion of the subordinate members of each node as the y-axis (the lower panel of Fig.\ref{fig:dendrogram}). It can visualize the astrophysical properties corresponding to branch nodes. So we can directly use 1000 km~s$^{-1}$, the typical galaxy cluster velocity dispersion, as a threshold to extract  physical structures from the dendrogram.

\subsection{Procedure}
Most mainstream data analysis languages provide related libraries and functions for hierarchical clustering, such as  ALGLIB\footnote{\url{https://www.alglib.net/dataanalysis/clustering.php}} of C$++$, Weka\footnote{\url{https://www.cs.waikato.ac.nz/ml/weka/}} and Hac\footnote{\url{http://sape.inf.usi.ch/hac}} of Java, scipy.cluster.hierarchy and sklearn.cluster  \footnote{\url{https://scikit-learn.org/stable/modules/generated/sklearn.cluster.AgglomerativeClustering.html}} of Python, hclust of R, clusterdata of Matlab,  dendro\footnote{\url{https://github.com/low-sky/dendro}} of IDL, etc.

These codes all implement the core function of the algorithm, which can draw the hierarchical tree in different linkage methods. However, in practical applications the analysis and extraction of structures are closely associate with specific physical problems, a lot of additional development work is required on the basis of existing functions. There are specialized packages of hierarchical clustering in astronomy to deal with specific problems. For example, astrodendro \footnote{\url{http://ascl.net/1907.016}} -- clustering by radio flux of molecular cloud, acorns\footnote{\url{http://ascl.net/2003.003}} -- clustering by radial velocity of molecular cloud, AMANDA\footnote{\url{https://ascl.net/1503.006}} -- clustering by multiple correlation matrix, etc. In researches, we need to select appropriate tools according to our own data characteristics and research goals and modify them.

\section{Application}
\label{sec:appl}
The hierarchical clustering algorithm has a fairly wide applicability because it does not rely on any priori information or models. It has been applied in many fields of astronomy. In the following we review the history and current status of the application of hierarchical clustering methods in different fields.

\subsection{Analysis of The Structure of Celestial Bodies}
All levels of objects in the universe are clustered by gravity and merge to evolve the hierarchical cosmic structure we see today. In principle, all systems sustained by gravity can be analyzed by hierarchical clustering and presented with a dendrogram.

\subsubsection{Galaxy Clusters}
\label{label:gcluster}

As the largest gravitationally-bound system in the universe, galaxy clusters play an important role in the evolution scenario of the universe. On the one hand, they are joints of the large-scale structure; on the other hand, they provide environments for galaxy formation and evolution. The search for galaxy clusters is an important topic. The hierarchical clustering was first introduced to the study of astronomy by \cite{1978A&A....63..401M} to analyze nearby groups and clusters of galaxies. He designed a dimensionless distance with spatial positions and velocities of galaxies as the metric to explore nearby groups in the Leo region. This attempt made the hierarchical clustering the first astronomical three-dimensional group analysis method. It has attracted the attention of early redshift survey researchers. \cite{Tully1980} tested this algorithm. He designed a quantity related to gravitational force (luminosity divided by the square of the spatial distance) as the distance to analyze the field of galaxy group NGC 1023. In 1987, he designed a density-like metrics to recognize galaxy clusters in the Nearby Galaxies Catalog \citep{Tully1987}. \cite{Gourgoulhon1992} used a hierarchical clustering algorithm to compile a catalog of all-day galaxy clusters within 80 Mpc of the Milky Way.

Meanwhile, \cite{1982Huchra} designed a method for processing single linkage results with a fixed threshold to analyze galaxy redshift survey data, as known as the Friend-of-Friend(FoF) algorithm. It corresponds to a fixed level in the hierarchical clustering dendrogram, thus simplifying the computational process and the complexity of operations. Hierarchical clustering and FoF were main group finders at that time. \cite{1992Garcia} systematically compared the differences in galaxy clusters obtained by the two methods. However, due to the lack of an objective and accurate evaluation standard, it is difficult to say which method is more reliable. \cite{1993Garcia} created a galaxy cluster catalog by combining these two methods. Later, FoF algorithm has gradually become a common tool in the astronomical community because of its speed and convenience, and is widely used.

However, there are still researchers who have been persistently exploring the potential of hierarchical clustering for astronomical structure identification. \cite{1996A&A...309...65S} found when using intergalactic binding energy as a metric, (see equation\ref{eq:pairwise-energy}), the substructure within galaxy clusters can be visualized on a dendrogram. The typical calculation equation of binding energy as follows:
\begin{equation}
E_{ij}=-G\frac{m_i m_j}{R_p}+\frac{1}{2}\frac{m_i m_j}{m_i+m_j}\Pi^2 \;,
\label{eq:pairwise-energy}
\end{equation}
where G is the gravitational constant, $m$ is the star mass, $R_p$ is the pairwise relative distance, $\Pi$ is the velocity difference based on spectral redshifts. This is an important breakthrough compared to previous work. As a quantity with a clear physical meaning, binding energy organically combines spatial location and velocity in the data, not only does it solve the problem of scaling the definition of "distance", but the resulting dendrogram is also clearly interpretable. 
\cite{1996A&A...309...65S} also found, the dendrogram obtained by single linkage gives the most reasonable results for galaxy clusters detection. There are other work \citep[eg.][]{Adami2005,2014A&A...561A.112G} adopting their method to analyze substructures of clusters. However, their work did not provide a standardized way to trim trees.

\begin{figure}[htp]
\centering
\includegraphics[width=0.48\textwidth]{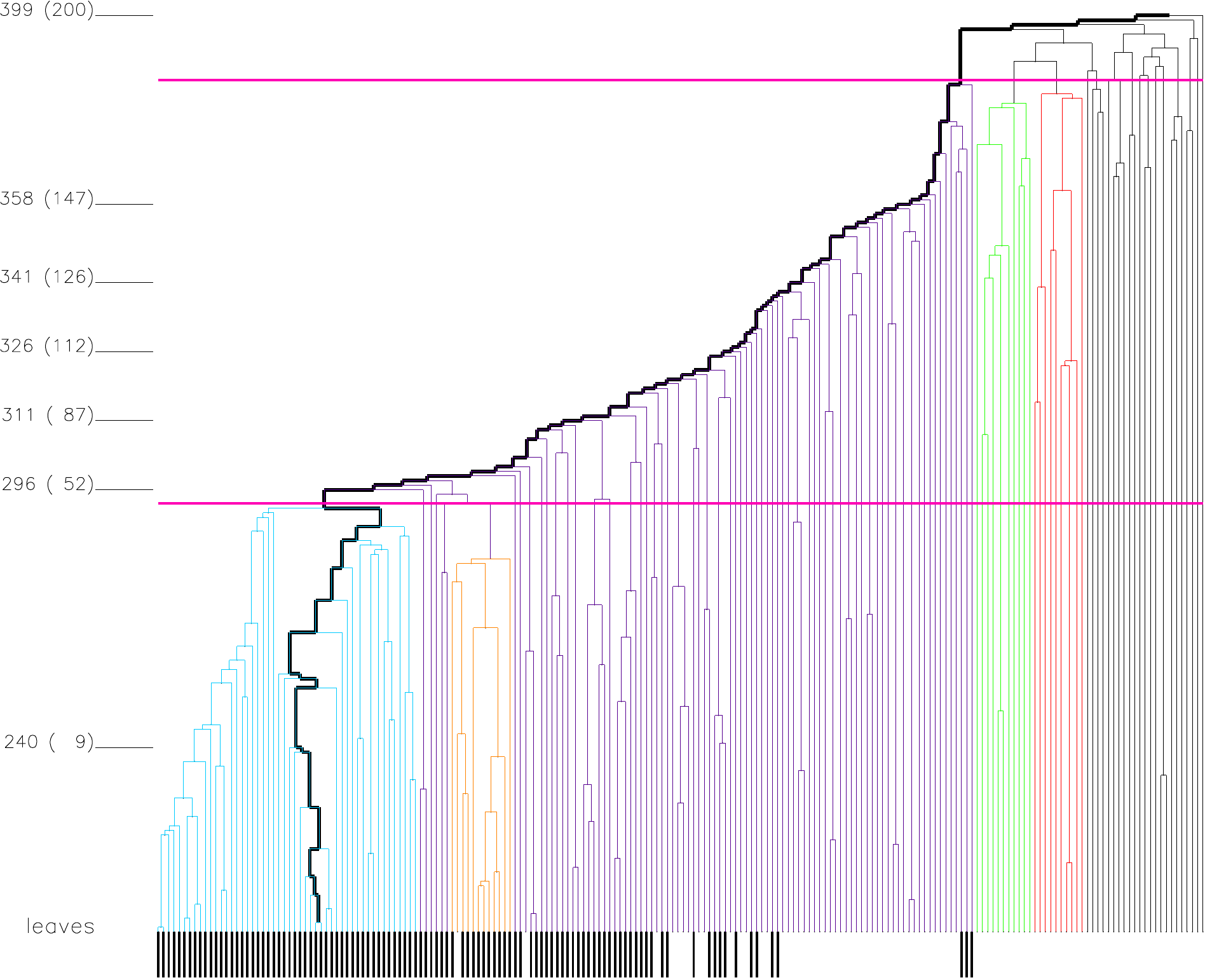}
\includegraphics[width=0.48\textwidth]{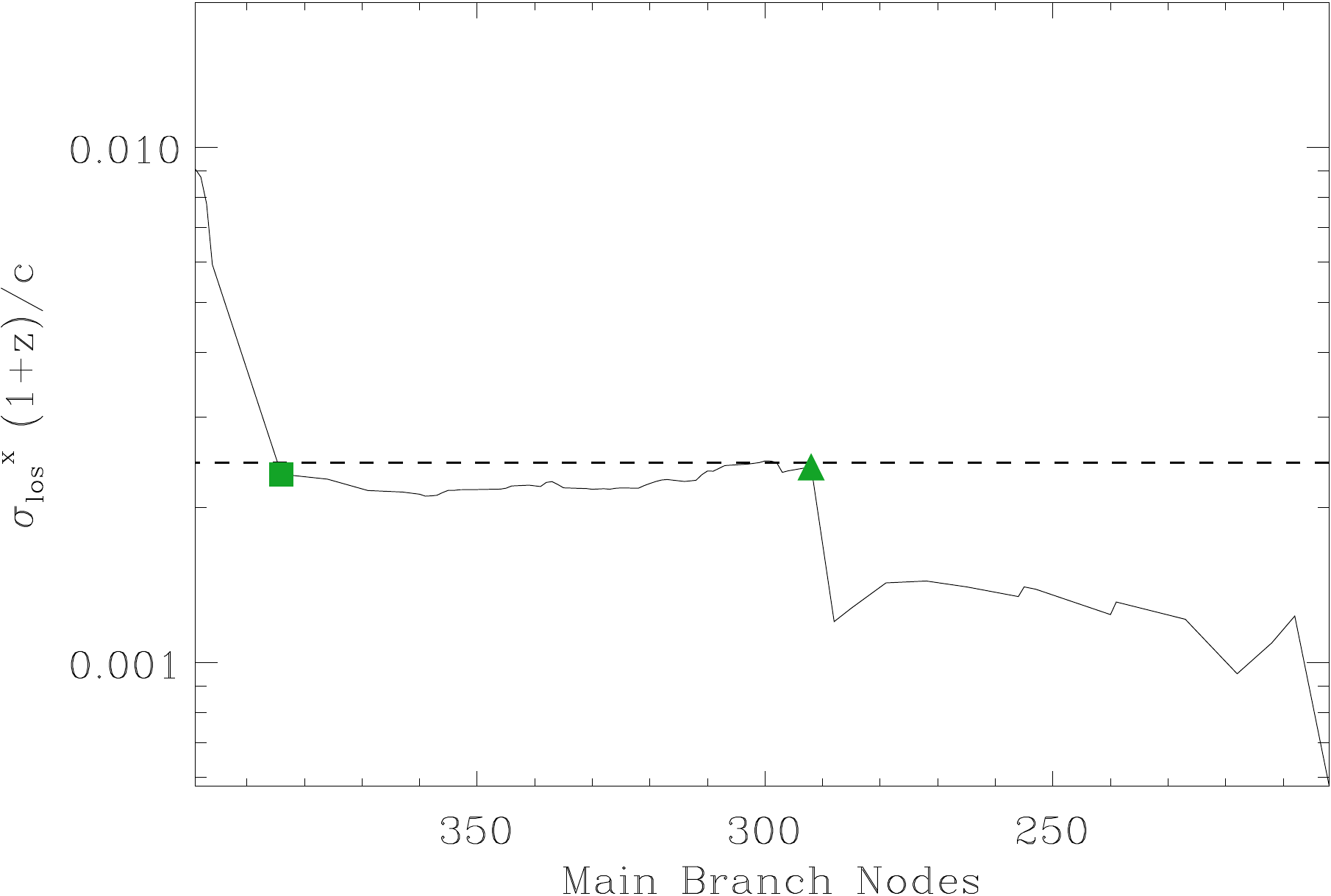}
\caption{The upper panel shows a dendrogram of a simulated galaxy cluster, with the main branches depicted by a solid black line. The lower panel shows the velocity dispersion contour corresponding to this branch. The green dots at the ends of the velocity dispersion platform are the two thresholds for identifying galaxy clusters and substructures, respectively, corresponding to the two horizontal lines in the left panel. These figures are taken from \cite{2013ApJ...768..116S}, reproduced with permission.} %\cnenfigcaption
\label{fig:plateau}
\end{figure}

\cite{Diaferio1999} further proposed that the velocity dispersion plateau in the dendrogram could be used to identify the galaxy clusters contained therein. The specific approach is: from the root node, every step moves along the branch with more members, then the main branch of the dendrogram can be obtained. When there is a dominant galaxy cluster in the field, velocity dispersion ($\sigma$) from nodes on the main branch will from an obvious plateau as shown in the Fig.\ref{fig:plateau}. According to this plateau, the galaxy cluster can be identified in the dendrogram. This procedure provides a quantitative operational standards of structure extraction in dendrogram. \cite{2013ApJ...768..116S} systematically tested this method by using data from large scale cosmological simulations, which confirmed the reliability of it in probing galaxy cluster members.

In fact, this velocity dispersion plateau can be used to detect not only members of galaxy clusters, but also substructures in galaxy clusters. \cite{2015ApJ...810...37Y} confirmed this idea through a systematic test of its capabilities in detecting the structure of galaxy clusters using numerical cosmological simulation data. In 2016, they applied this method in galaxy cluster A85, finding an optical substructure moving in the direction of the line of sight in its center, which can well explain the spectral shift of the X-ray gas at the core \citep{Yu2016}. \cite{2018ApJ...863..102L} also applied it to galaxy clusters A2142, confirmed that the substructure given by this method is consistent with the detection by other methods such as X-ray images and weak gravitational lensing.

\begin{figure}
\centering
\includegraphics[width=0.48\textwidth,trim=0 1 0 0]{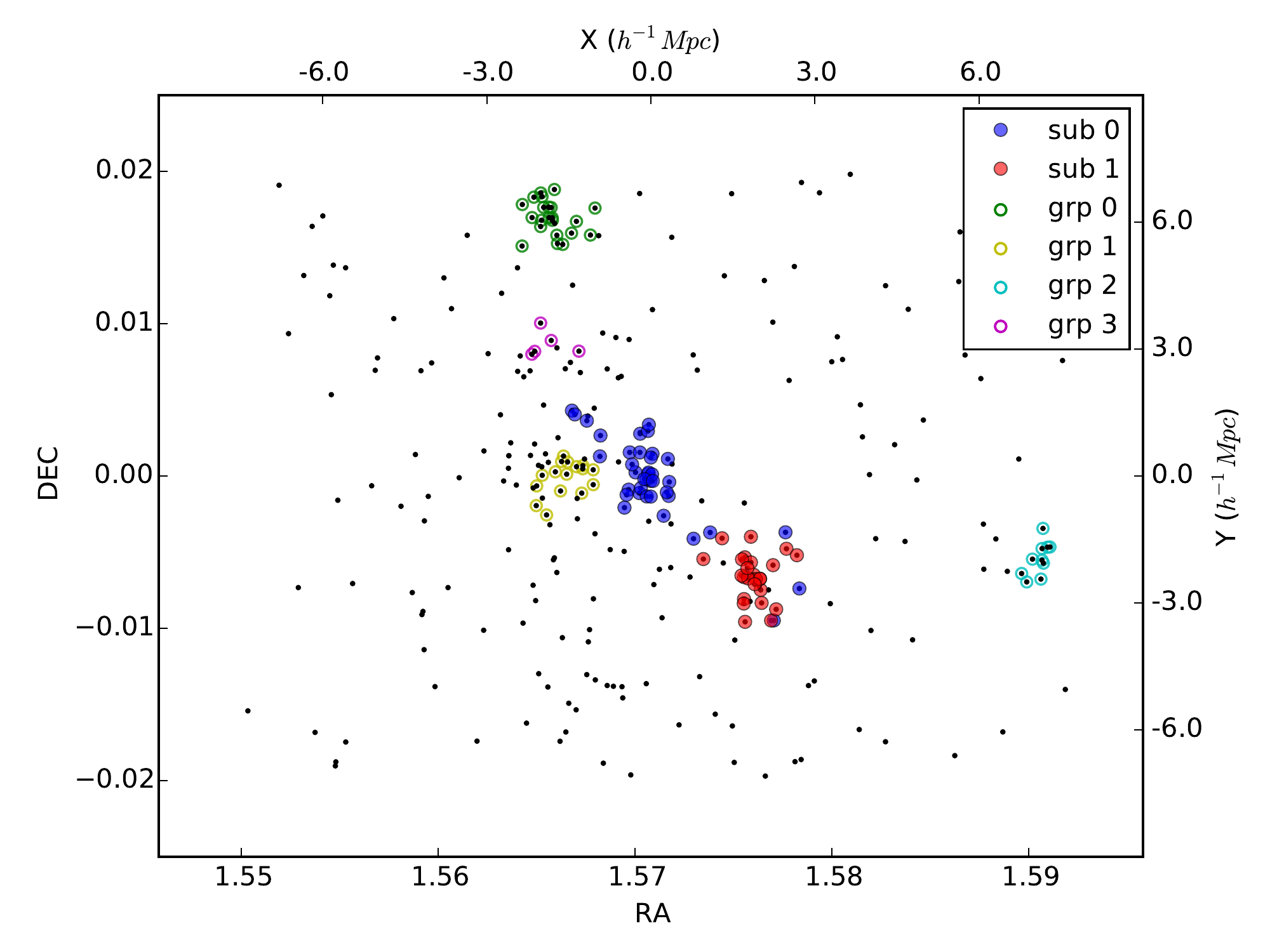}
\includegraphics[width=0.49\textwidth,trim=0 0 0 1]{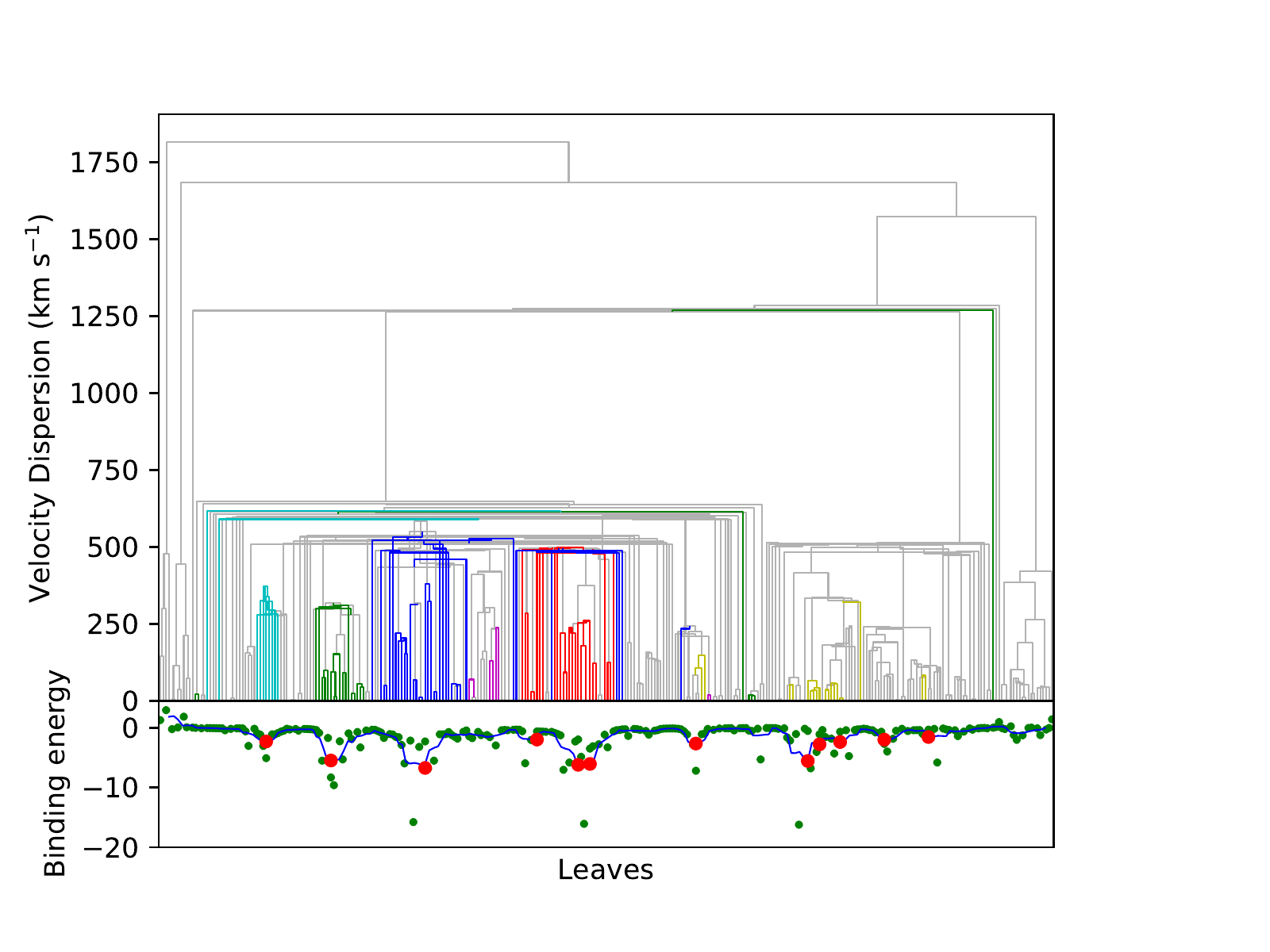}
\caption{The upper panel shows the real distribution of galaxy cluster members in the cosmological simulation data. The different colors represent the members of different clusters. The lower panel shows their corresponding dendrogram, with the galaxy position on the horizontal axis and the velocity dispersion of each node on the vertical axis. The galaxies in the field of view are connected successively according to their projected binding energy. It is clear that members of galaxy clusters gather in the dendrogram with different velocity dispersion platforms. These figures are from \cite{Yu2018b}. }
\label{fig1}
\end{figure}

However, the approach of trimming the dendrogram according to the velocity dispersion plateau is mainly designed for individual galaxy clusters and is not suitable for large field-of-view multi-target studies. Fig.\ref{fig1} shows the spatial distribution of the simulated galaxy data and their corresponding dendrograms. Galaxy clusters with different masses have different velocity dispersion platforms and are hard to be extracted with a uniform threshold. \cite{Yu2018b} proposed a new algorithm to trim the dendrogram----“Blooming Tree”. Instead of setting a single threshold on the velocity dispersion, the new method traces all leaves and takes multiple properties ---- the velocity dispersion, membership, and spatial distribution of each node into account to decide whether to trim the branch. The Blooming Tree algorithm is able to automatically detect all dense areas in the field without specifying their positions or redshifts. It significantly reduces the manual intervention in the analysis, improves the applicability of the method, and make the algorithm possible to analyze large-scale structures of the universe.

\subsubsection{Supercluster}
48
Superclusters are supermassive objects formed by the gravitational aggregation of multiple galaxy clusters. Because their scales are larger than 10 Mpc, the relaxation time exceeds the age of the universe. Therefore no distinct mass center or boundary has been formed. This poses a great difficulty for its definition and search. The methods of detecting superclusters usually derived from the algorithm of detecting galaxy clusters, the main methods are density of galaxy numbers\citep{2007A&A...462..811E,2012A&A...539A..80L}, Friends-of-Firends algorithm\citep{2011MNRAS.415..964L,2013Chon,2014Chow,2021Liu}, etc. However, our knowledge of such non-relaxation systems is still very limited because of the lack of effective evaluation methods.

\begin{figure}[htp]
\centering
\includegraphics[width=0.49\textwidth]{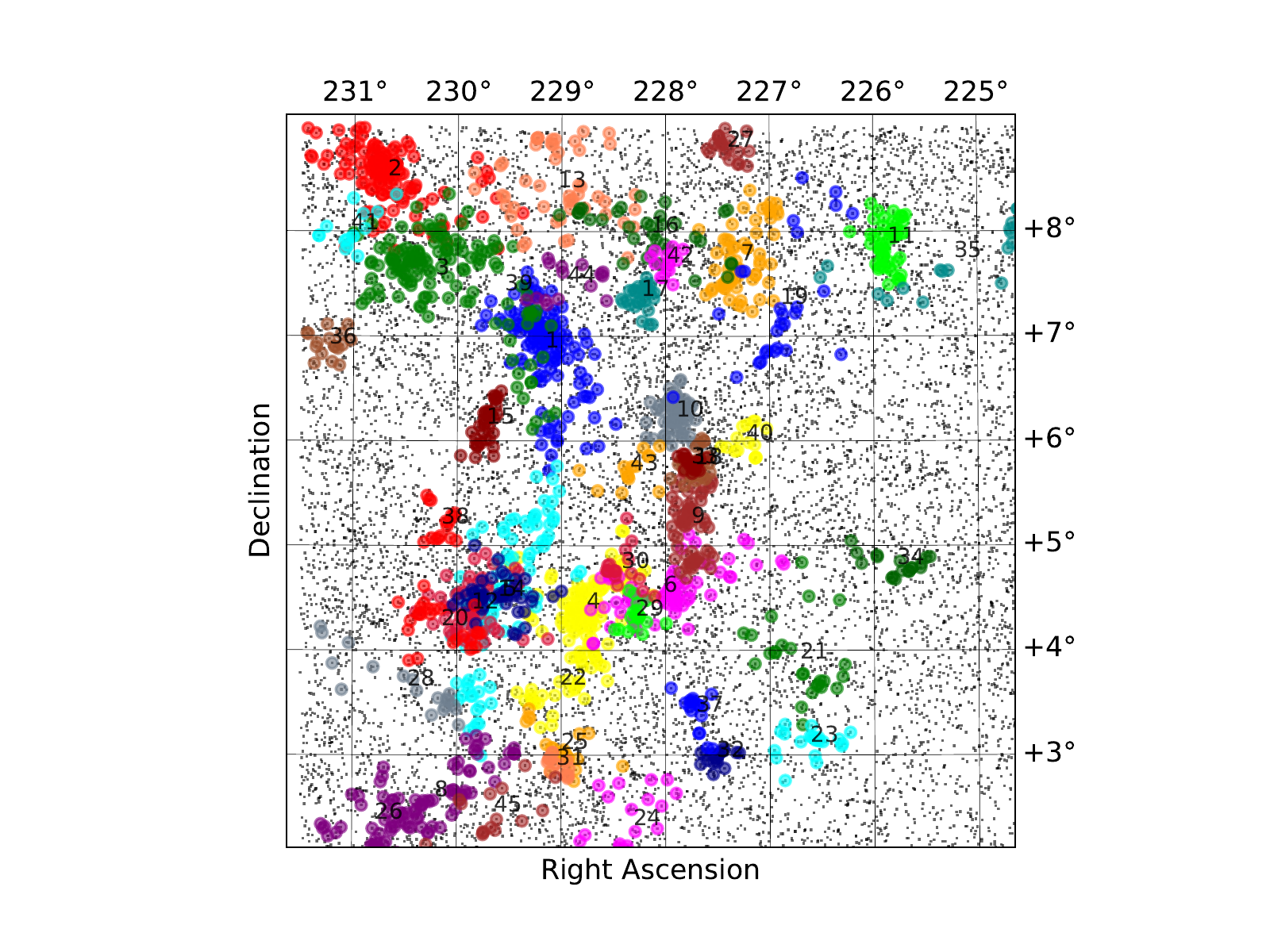}
\includegraphics[width=0.49\textwidth]{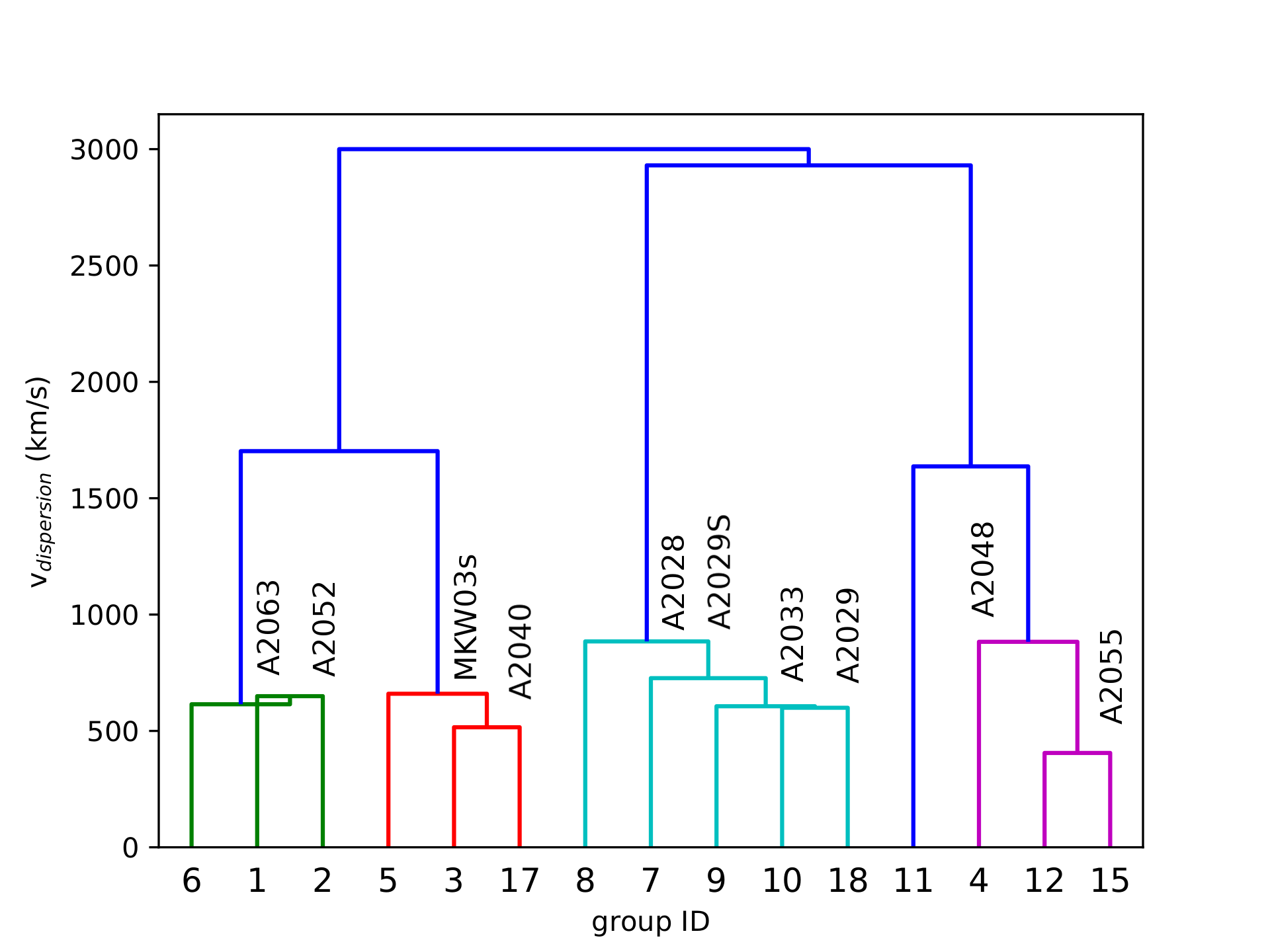}
\caption{The top panel shows the spatial distribution of galaxy clusters/groups detected by the blooming tree algorithm in the sky region near supercluster A2029. The different systems are distinguished by different colors and the numbers are located at the center of each cluster. A simplified dendrogram of these structures is shown on the bottom, with names of known galaxy clusters labelled above the nodes.}
\label{fig:super}
\end{figure}

Hierarchical clustering identifies galaxy clusters while also giving the hierarchy of their surroundings, and thus can provide new ideas for the detection of superclusters. We apply this algorithm (Blooming Tree) to the spectroscopic data of Sloan Digital Sky Survey (SDSS), and find many galaxy cluster candidates at different redshifts in the field of view of galaxy cluster A2029, as shown in Fig.\ref{fig:super}. The subordination between these structures can be seen directly in the simplified dendrogram on the bottom. 
Based on the typical velocity dispersion of the relaxation system at each level in the universe given by the virial theorem, we can roughly determine the position of the system at each level in the dendrogram. Velocity Dispersion 500km/s corresponding to galaxy groups, 1000km/s corresponding to galaxy clusters, the platform at 2000km/s indicates the presence of superclusters. Systems located in the same branch have strong gravitational associations with each other, while structures between different branches are more likely to have no direct gravitational connection. Furthermore, the order in which the systems merge into the branches can also tell how closely the members are related, so that the fiber structure can be found (We have detected in our analysis of galaxy cluster A85 an optical galaxy population corresponding to its outer X-ray fiber structure \citep{Yu2016}), even the future merging sequence could be predicted. \cite{2020A&A...637A..31S} had also introduced hierarchical clustering algorithms to the detection of fiber structures in superclusters of galaxies. %It is reasonable to expect that hierarchical clustering algorithms will play a greater role in this direction in the future.

\subsubsection{Star Cluster}
Stars may also assemble into clusters like galaxies. There are two types of star clusters: globular clusters and open clusters. The former are mostly located in galactic halos and have a regular spherical shape with a high density of member stars, and their origin is not clear; while the latter are born in molecular clouds and slowly disintegrate under many effects such as environmental perturbations, galactic tidal forces, and stellar evaporation, so their shape is very irregular. Star clusters are important objects for studying the evolution of stars and probing the structure of galaxies. The identification of cluster members is usually based on two a priori assumptions: one is the spatial position and motion information of the stars, assuming that cluster members have the same kinematic characteristics; the other is the magnitude and color information, assuming that member stars have the same evolutionary trends. Researchers have introduced various algorithms to identify cluster members based on these assumptions.
Back in the middle of the last century, \cite{1958AJ.....63..387V} and \cite{1971A&A....14..226S} had introduced the maximum likelihood method and double Gaussian fit to find the members of clusters. Many researchers have conducted numerous follow-up studies based on it \citep{1990A&A...237...54Z,2006A&A...446..949D,2010A&A...516A...3K,2014A&A...563A..45S,2016MNRAS.457.3949S}, and this method is still the mainstream method in this field. In order to avoid the prior error introduced by Gaussian distribution, some researchers have introduced nonparametric methods to perform the fitting. However, these methods cannot get rid of the reliance on priori information. Recently, more modern clustering methods have been introduced into the field. \cite{2011Schmeja} tested four clustering methods with two-dimensional projection data in 2011: star counts, nearest neighbour, Voronoi tessellation, minimum spanning tree(MST), etc, which was found to have the best combined performance for nearest-neighbor particles. \cite{2014A&A...561A..57K} developed UPMASK method, combining with principal component analysis(PCA) and k-means algorithm, and applied for cluster analysis with the Gaia DR2 data. In addition, the effect of the DBSCAN algorithm was also tested by \cite{2014RAA....14..159G} and \cite{2018A&A...618A..59C}, respectively. The key parameters of these machine learning methods generally lack a reasonable physical interpretation, which has hindered the application and popularity of these methods.
\begin{figure}[htp]
\centering
\includegraphics[width=0.49\textwidth]{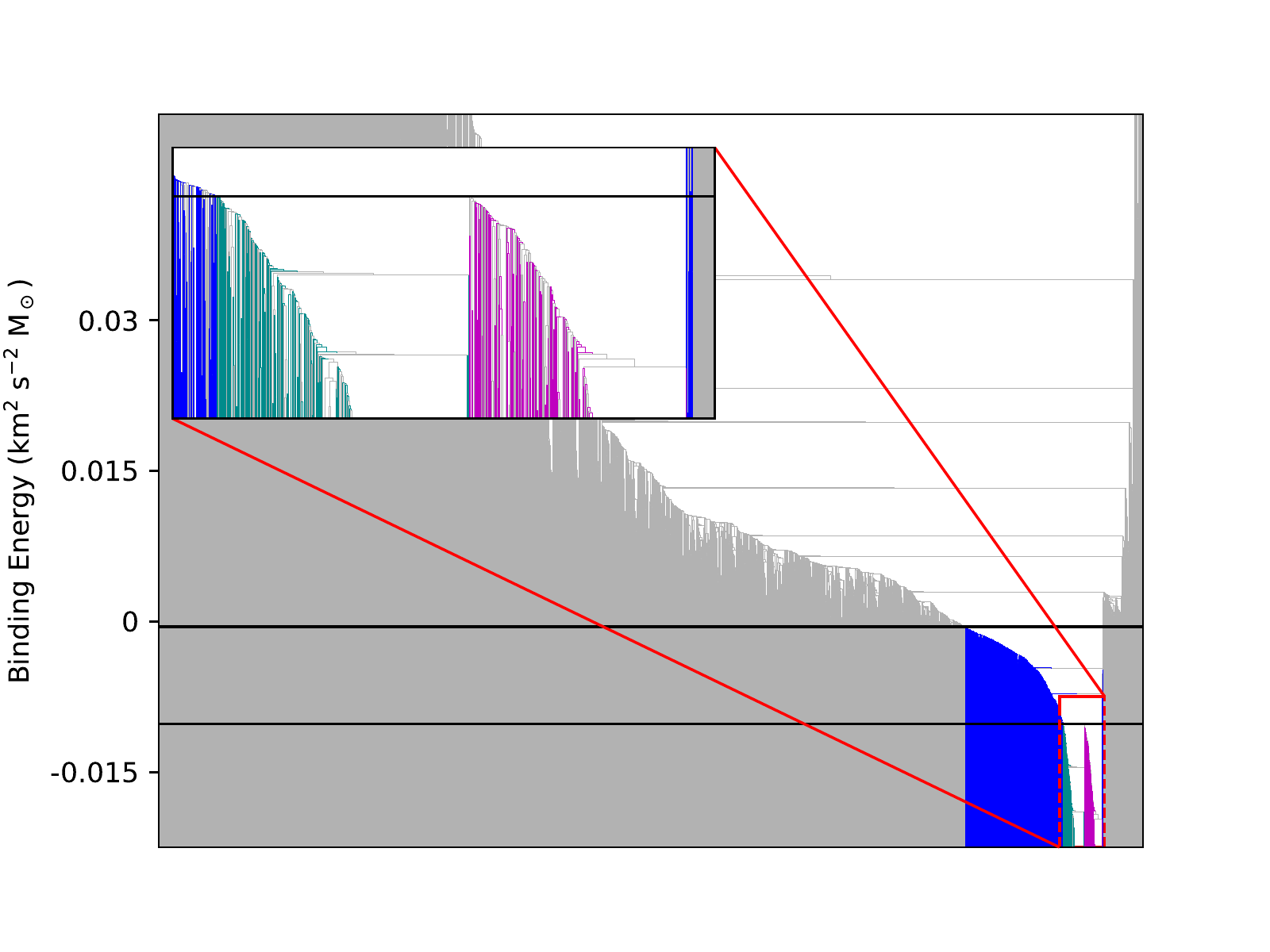}
\includegraphics[width=0.49\textwidth]{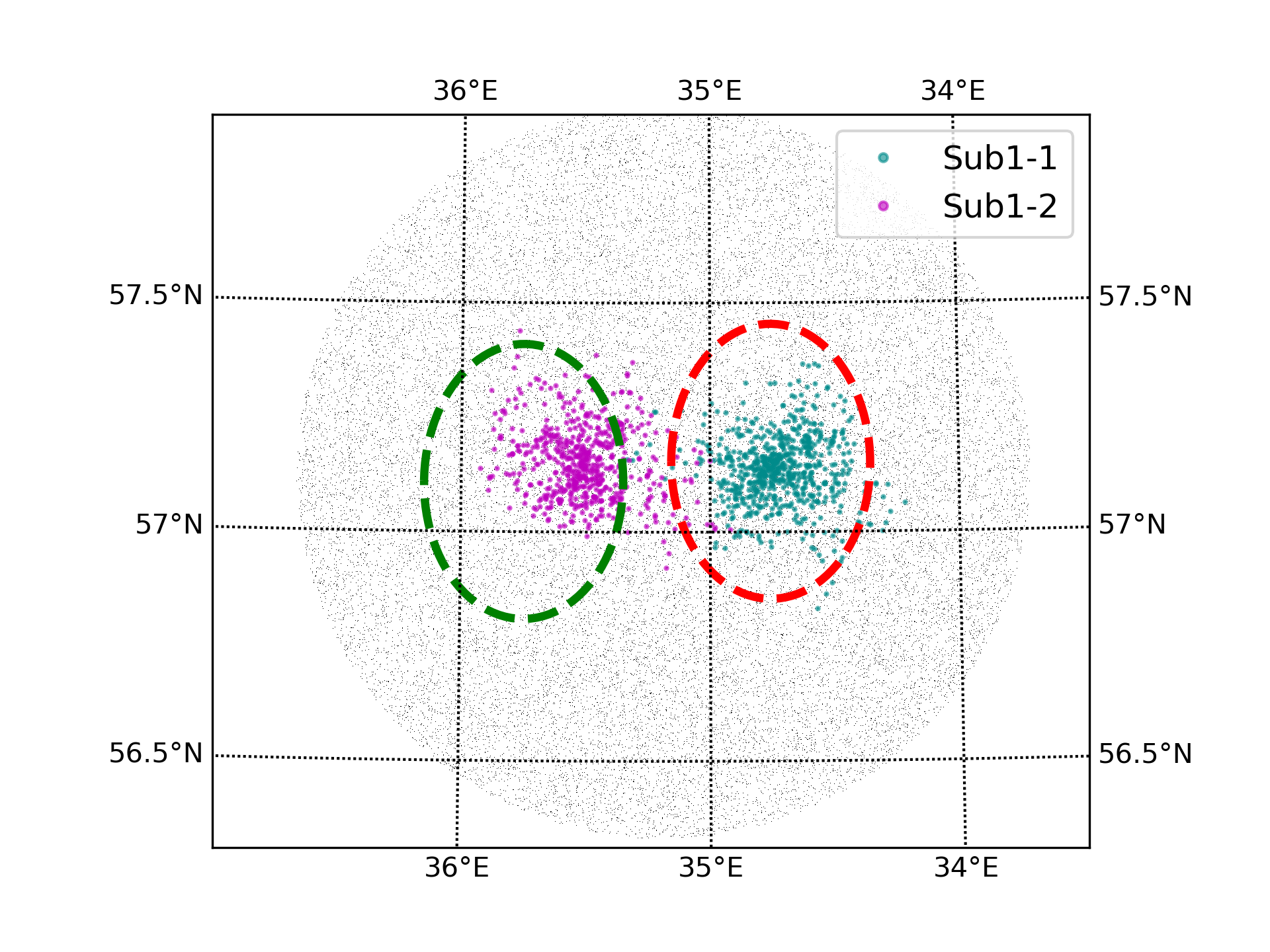}
\caption{The upper panel shows the dendrogram of the Perseus Double Cluster. The horizontal axis is the binding relation of 32672 stars in the field of view. The vertical axis is the binding energy of each node. The blue color is the member of the gravitational binding system sub1 (binding energy less than 0). Cyan and magenta stars correspond to the two subclusters respectively. The black horizontal lines are the two thresholds to identify structures. The lower panel shows the spatial distribution of the two subclusters identified by the algorithm. The elliptical dashed line is the reference range of subclusters. These figures are from \cite{2020Yu}.}
\label{fig:perseus}
\end{figure}

Hierarchical clustering methods do not require prior assumptions about the morphology or distribution of clusters. This is advantageous for exploring the boundaries of clusters and detecting clusters with low density or even disintegrating. Moreover, hierarchical clustering methods can not only give information about specific members, but also provide the hierarchical relationship of these members in the gravitational potential well. In fact, \cite{1983Rood} had tried using hierarchical clustering algorithm  to analyze nearby stars, but no clear conclusions were obtained. \cite{2020Yu} used Gaia DR2 data, tested the potential of hierarchical clustering algorithms for cluster structure analysis with the example of the open cluster Perseus Double Cluster. Perseus Double Cluster is a young cluster located at the distance of 2.34 kpc from the Sun,  which contains two subclusters separated by about 46 arc minutes. It is not clear how the two subclusters overlap each other. \cite{2020Yu} selected the spatial position and proper motion in the field of view, with projection binding energy as a metric. The dendrogram obtained is shown in the Fig.\ref{fig:perseus}. The subsidence caused by the gravitational potential well of the cluster can be clearly seen in the dendrogram with the binding energy as the y-axis, and the stars deep in the potential well belong to two sub-clusters. According to the thresholds obtained from the velocity dispersion platform method, the halo members of the double cluster and the members of the two subclusters can be extracted from the field full of stars. The cluster members given by hierarchical clustering show a fairly good agreement in terms of color, parallax, etc. Compared with the traditional cluster member identification methods, the hierarchical clustering method only considers the kinematic information of the stars and does not depend on their physical properties such as luminosity or color. It provides a new way for the identification and study of cluster members.

\subsubsection{Molecular Cloud}

Although molecular clouds are not as discrete in distribution as stars or galaxies, they also have hierarchical features. Many studies have shown that high-density features in molecular clouds have relatively small physical scales and are always located in environments with low-density gas; for any scale, dense structures on small scales are more numerous than sparse structures on large scales \citep{2008ApJ...679.1338R}. Dendrogram using hierarchical clustering algorithms can naturally identify such nested features and thus reflect the relationships between different types of structures in the data.

\begin{figure}[htp]
\centering
\includegraphics[width=0.32\textwidth,trim={5cm 0 5cm 0}]{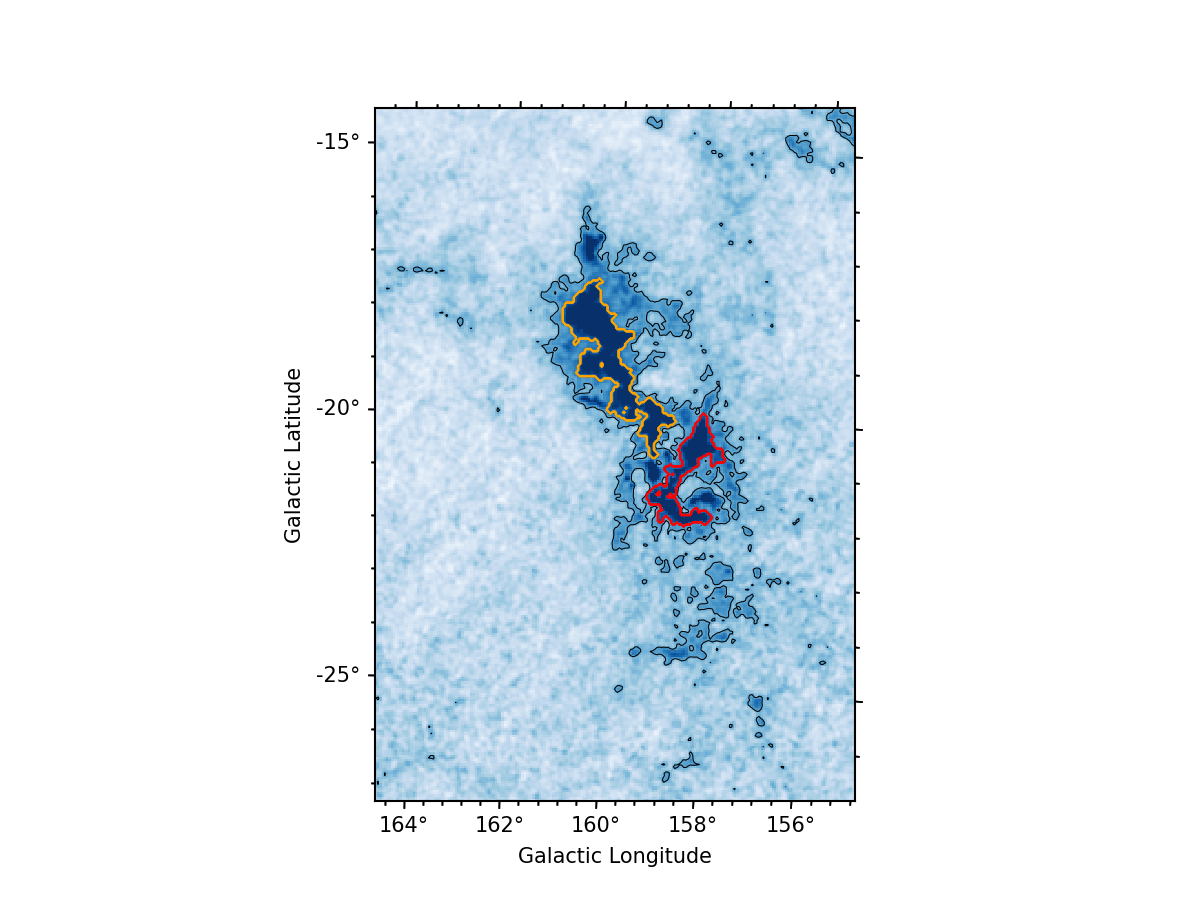}
\includegraphics[width=0.48\textwidth]{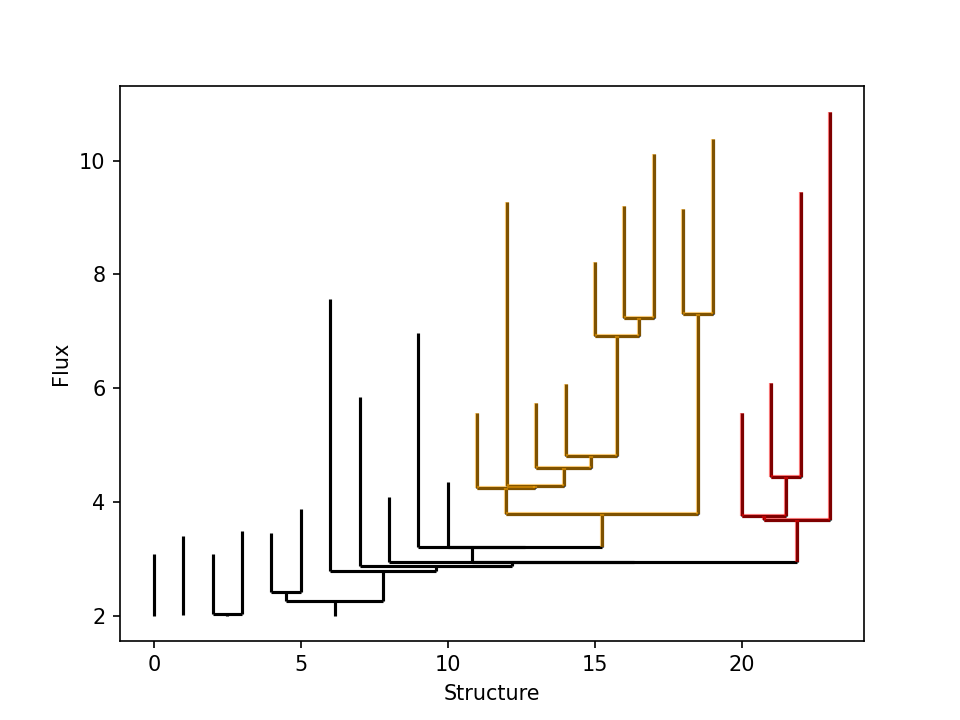}
\caption{The top panel shows the image of a molecular cloud with the Galactic coordinate. The bottom panel shows the dendrogram corresponding to this region, with the surface brightness(K) of each pixel point on the y-axis, characterizing the emission intensity. The red branches in the dendrogram correspond to the region outlined in red in the left panel, while the orange branches are from the orange region. These figures are made with the Python module astrodendro.}
\label{fig:quad1}
\end{figure}

\cite{1992ApJ...393..172H} introduced the hierarchical clustering method to the study of the internal structure of molecular clouds for the first time. The dendrogram based on the molecular cloud column density and applied to the Taurus complex reveals a hierarchical structure in the star-forming molecular cloud complex. \cite{2008ApJ...679.1338R} take each pixel point as a unit and use the position-position-velocity (PPV) data of that pixel point as a parameter in hierarchical clustering to calculate its correlation with surrounding pixel points. By quantifying the self-gravitating (self-gravitating) bound states of molecular clouds, they are able to distinguish between independent molecular clouds and substructures within molecular clouds. The new method was applied to the self-gravitational analysis and the identification of giant molecular clouds of Perseus L1448. It was compared with CLUMPFIND, the dominant algorithm for studying the structure of giant molecular clouds at that time. CLUMPFIND based on FoF algorithm, connects based on the values of the surrounding pixels and requires segmentation of the molecular cloud, focusing more on small-scale structures. Dendrogram, however,  avoids segmentation and is able to cover large scale structures. Their results were later published in Nature letter \citep{2009Goodman}. Then, hierarchical clustering gained recognition in the field of molecular cloud research, and many subsequent studies were carried out based on it.

\cite{2015MNRAS.454.2067C} designed the SCIMES algorithm based on the \cite{2008ApJ...679.1338R} method, using PPV, 3D PPP data or 2D PP images as input data, and testing with data from the Orion-Chiron complex. They found characteristic volume and integrated CO luminosity can be used as valid criteria for trimming dendrograms to extract structure. And SCIMES performs better than CLUMPFIND in complex environments and high resolution data.
\cite{2016ApJ...822...52R} used $^{12}$CO CfA-Chile survey data, dividing the galactic plane into four quadrants according to the size of the longitude, and then plotted dendrograms, respectively. Where the first quadrant (${13^\circ<l<17^\circ}$) has a rich molecular cloud structure and a more complete research history, the dendrogram performs well in this complex environment, which is shown in Fig.\ref{fig:quad1}; and they obtained a catalog of 1064 giant molecular clouds in the entire galactic plane. \cite{2020MNRAS.493..351C} applied the \cite{2008ApJ...679.1338R} method to three-dimensional dust extinction and stellar color residual data to derive a distance catalog for giant molecular clouds at low latitudes, within 4 kpc of the Sun, with much higher distance accuracy than other methods. This is the first catalog of giant molecular clouds to derive distances in a direct way. Similarly, \cite{2022MNRAS.511.2302G} applies hierarchical clustering algorithm to the three-dimensional extinction map of the southern sky of the Milky Way, systematically verify and analyze the molecular clouds in there, and have obtained the first large-scale molecular cloud table of the southern sky with accurate distance measurement.

In addition to the usual PPV data, more multidimensional data can also be added for hierarchical clustering analysis. \cite{2019MNRAS.485.2457H} studied the internal dynamics of G0.253+0.016, one of the most massive and dense molecular clouds at the center of the galaxy. They added the centroid velocity and velocity dispersion to the spatial information, and found that the internal dynamical state is complex and has a hierarchical structure. \cite{2020MNRAS.497.4517S} used magnetohydrodynamically simulated three-dimensional density fields to link the dendrogram structures through time, and by imposing constraints on the background cutoff, the minimum density increase for creating new structures, the minimum structure size, and the connection distance, they draw a physically consistent dendrogram and compare it with observations to study the time evolution of all dense cores in the region. In addition to the direct clustering of molecular clouds, the structure of molecular clouds can be studied indirectly by clustering of stars. \cite{2019A&A...630A.137G} used Gaia DR2 data to cluster the newly born stars in the Taurus complex, linking the stellar positions and physical parameters to the molecular clouds in which they are located. They grouped the molecular clouds corresponding to these stars in space into 21 subgroups, revealing the coupling between stars and molecular clouds in velocity and space.

From these works, it can be seen that the hierarchical clustering method has become increasingly widespread and diverse in the application of molecular cloud research.

\subsection{Classification of Celestial Bodies}
The classification of celestial bodies is a fundamental part of astronomical research. Traditionally, the classification of celestial bodies is done manually by astronomers based on experience. As astronomy technology advances, astronomical data sets become larger and larger. To analyze these data effectively, discover previously unknown objects and features, and search for potential associations and patterns are new challenges posed by big data. The unsupervised hierarchical clustering method is particularly well suited to handle the need for automatic classification without a priori information. The hierarchical results it gives also provide valuable clues for the subsequent parameter tuning. Therefore there are many astronomical classification studies that have applied the hierarchical clustering method.

\cite{1987Eigenson} had tried applying hierarchical clustering algorithm to automatically classify globular and open clusters based on information such as luminosity, scale, color index, and metal abundance. This work has not received enough attention, though.

\cite{1990Zappala,1994Zappala} used this method to classify asteroid according to three-dimensional proper element. In 1995, they compared the results of this method with those of wavelet analysis using a sample containing 12,487 asteroids data, and found that the two methods gave essentially the same classification results for the major asteroid families \citep{1995Icar..116..291Z}. \cite{2007Carruba} referred to this method proposed that to hierarchical clustering by using proper frequency, which also identified asteroid families successfully. After optimized the distance metric, it even can recognize tynamical evolution of asteroid families \citep{2009Carruba}. They analyzed multiple asteroid families with this method \citep{2015Carruba,2016Carruba}. With the rapid increase in the number of asteroids, \cite{2014Milani} also used the hierarchical clustering to classify asteroid families, and even identify the collision fragments in them \citep{2019Milani}. A relevant review in this direction can be found in the paper published by \cite{2016IAUS..318...16K}.

Raw images or spectral data of celestial objects are usually not directly usable for drawing dendrograms and need to be quantified with the help of suitable methods.
\cite{2007Hojnacki} attempted an automated classification study of X-ray astronomical spectra. They first extracted the main information in the spectra by principal component analysis (PCA), then grouped them by the hierarchical clustering, and finally determined the members of each group by the k-means method.
In the same period, \cite{2007ApJ...666..475M} also used a similar idea to classify exoplanets. They first used PCA to identify key variables and then used these new variables to perform hierarchical clustering to obtain exoplanet classification. \cite{2009Marchi} also used this method to discuss the origin of exoplanets. \cite{2016Peth} used hierarchical clustering to process the results of PCA analysis of high-redshift optical galaxies to morphologically classify them in order to avoid the subjective errors introduced by manual classification. \cite{2018Hocking} also proposed a similar idea to identify galaxy morphology. They first tried to quantify the spatial morphology of galaxies, and then clustered the quantified results using the hierarchical clustering. Their classification results are generally consistent compared to the manual annotation results from the Galaxy Zoo \citep{2018Hocking}.

\begin{figure}[htp]
\centering
\includegraphics[width=0.48\textwidth]{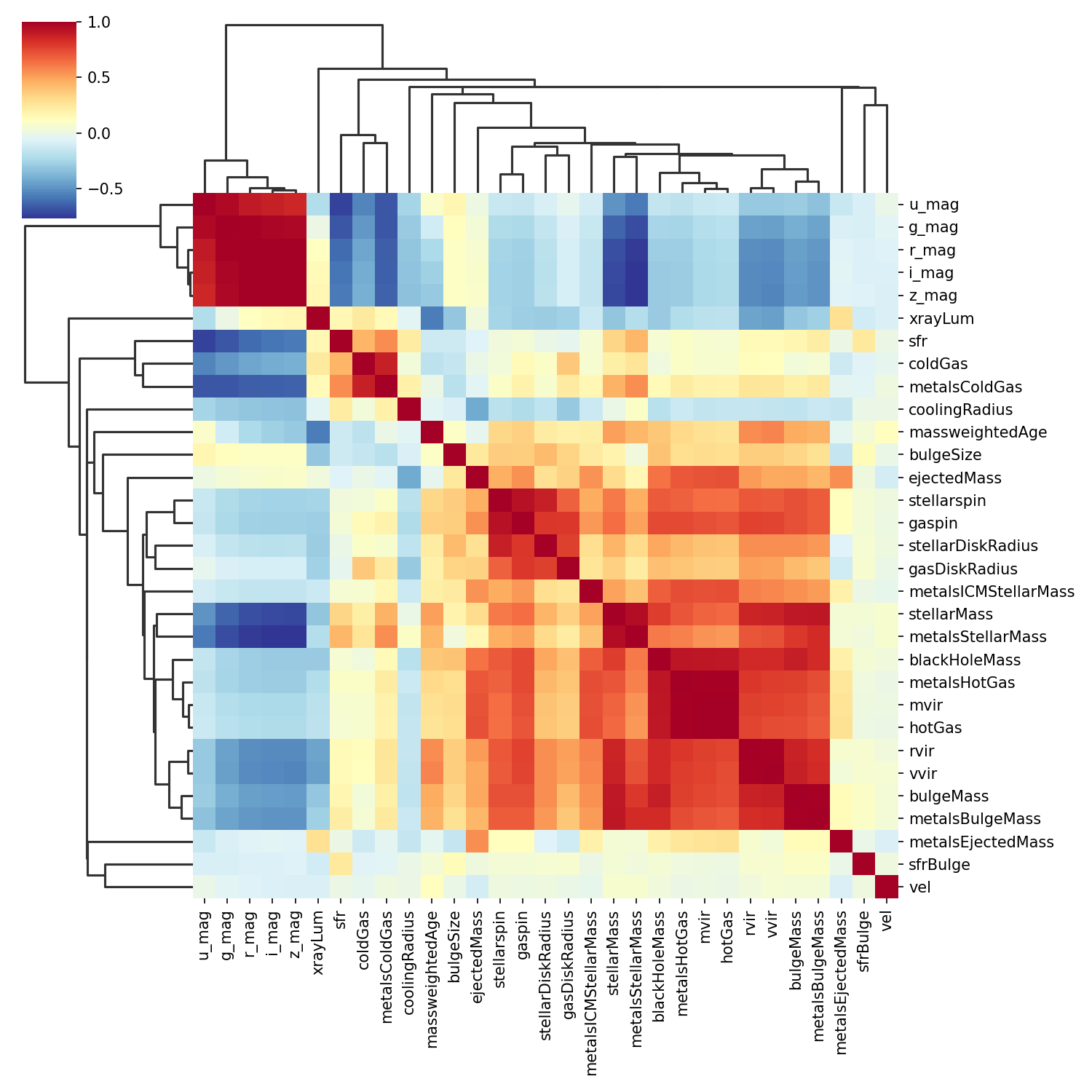}
\caption{Heat map of the correlation matrix of galaxy properties with the hierarchical clustering. Positive correlation in red, negative correlation in blue. It is clear that the ugriz magnitudes of galaxies are highly correlated. It is adapted from a figure of \cite{2015Souza} with the Seaborn module of Python.}
\label{fig:heatmap}
\end{figure}

Hierarchical clustering is often used in combination with correlation matrix or heat map, to reveal the correlation between multivariate variables. The early development of this method has been discussed detailed in  the review of \cite{2009:HCHM}. The popular statistical analysis language R and the popular programming language Python both have ready-made functions to plot heat maps. \cite{2015Souza} develops AMANDA in R to progress multidimensional astronomy data, which is able to hierarchical cluster correlation matrix. They test with simulated galaxy data, and Fig. \ref{fig:heatmap} shows that ugriz- band brightness in galaxies are highly correlated.
The first use of this technique in the field of astronomy may be \cite{2015Baron}. In their analysis of diffuse interstellar absorption bands (DIBs) of unknown origin, they used the hierarchical clustering method to process the correlation matrix between 142 DIB spectral lines to find potential families of spectral lines among them. They found that the spectral line families given by hierarchical clustering can be interpreted as absorption lines from the same molecule.

Hierarchical clustering has even been used for classification of time-varying data. \cite{2018Ma} tested this method in the type identification of coronal mass ejections (CMEs). However, they found that it was not as effective as the previous method, distance density clustering (DDC).

Overall, the hierarchical clustering algorithm is still in the early stages of experimentation in the field of celestial body classification except for asteroid identification. Although it has attracted the attention of many researchers in different astronomical fields, the procedure and the performance is different from case to case. To use it as a reliable automatic classification tool, a lot of tuning work is neccessary.

\section{Conclusion}

As a mature data analysis method, hierarchical clustering has attracted the attention of researchers in many branches of astronomy, and achieved a certain extent of recognition. But it's not a grab-and-go tool, it needs targeted design based on he actual data characteristics and scientific objectives.

The dendrogram obtained by hierarchical clustering is particularly suitable for describing hierarchical structure of gravitational systems. So it has a wide range of applications in gravitational systems at all levels of the universe like molecular clouds, star clusters, galaxy clusters and superclusters. In addition to identifying the members of these systems, its unique advantages are reflecting the dynamical state and evolutionary trend of members. As long as 
researchers handle the explanation and stability of the hierarchical clustering, it will play a bigger role in these fields.

In celestial body identification and classification, the unsupervised automated classification represented by the hierarchical clustering algorithm is expected to help astronomers search for unknown types of celestial objects and astronomical events. However, we also note that the form of implementation and the way of the hierarchical clustering algorithm work in different domains varies greatly due to the different knowledge structures and backgrounds of researchers. Although these papers use the same core algorithm, there are many differences in their algorithmic formulations, working logic, analytical tools, and even terminology names. If these can learn from each other, it will undoubtedly help us to make full use of the algorithm.

Of course, there are still some limitations of the hierarchical clustering algorithm. The key one is the computational complexity. The computation and storage of the two-by-two distance matrix between sets is an essential step for the hierarchical clustering. This determines the complexity of the hierarchical clustering algorithm to be at least $O(N^2)$, and it's difficult to handle data sets of more than 100,000 records now. And the same type of FoF algorithm only needs to use a specific trim threshold as a criterion to reduce the distance matrix to a Boolean matrix and retain only the grouping information, thus reducing the computational difficulty and storage cost. Its complexity is $O(N {\rm log}N)$. Therefore, to apply the hierarchical clustering algorithm to large astronomical data sets, further optimization is needed to reduce its computational cost.

With the construction of a new generation of astronomical equipment, modern astronomy is entering an era of all-sky, all-wavelength, all-time observations, and the amount of data to be analyzed is exploding. New tools need to be developed and evolved to make truly valuable scientific discoveries with limited staff and computing power.

\printcredits

\bibliographystyle{cas-model2-names}

\bibliography{cas-refs}

\begin{thebibliography}{77}
\expandafter\ifx\csname natexlab\endcsname\relax\def\natexlab#1{#1}\fi
\providecommand{\url}[1]{\texttt{#1}}
\providecommand{\href}[2]{#2}
\providecommand{\path}[1]{#1}
\providecommand{\DOIprefix}{doi:}
\providecommand{\ArXivprefix}{arXiv:}
\providecommand{\URLprefix}{URL: }
\providecommand{\Pubmedprefix}{pmid:}
\providecommand{\doi}[1]{\href{http://dx.doi.org/#1}{\path{#1}}}
\providecommand{\Pubmed}[1]{\href{pmid:#1}{\path{#1}}}
\providecommand{\bibinfo}[2]{#2}
\ifx\xfnm\relax \def\xfnm[#1]{\unskip,\space#1}\fi
%Type = Article
\bibitem[{{Adami} et~al.(2005){Adami}, {Biviano}, {Durret} and
  {Mazure}}]{Adami2005}
\bibinfo{author}{{Adami}, C.}, \bibinfo{author}{{Biviano}, A.},
  \bibinfo{author}{{Durret}, F.}, \bibinfo{author}{{Mazure}, A.},
  \bibinfo{year}{2005}.
\newblock \bibinfo{title}{{The build-up of the Coma cluster by infalling
  substructures}}.
\newblock \bibinfo{journal}{Astronomy and Astrophysics} \bibinfo{volume}{443},
  \bibinfo{pages}{17--27}.
\newblock \DOIprefix\doi{10.1051/0004-6361:20053504},
  \href{http://arxiv.org/abs/astro-ph/0507542}{\tt arXiv:astro-ph/0507542}.
%Type = Book
\bibitem[{{Babu} and {Feigelson}(1996)}]{1996Babu}
\bibinfo{author}{{Babu}, G.J.}, \bibinfo{author}{{Feigelson}, E.D.},
  \bibinfo{year}{1996}.
\newblock \bibinfo{title}{{Astrostatistics}}.
\newblock \bibinfo{publisher}{CRC Press}, \bibinfo{address}{Boca Raton,
  Florida}.
%Type = Article
\bibitem[{{Baron}(2019)}]{2019Baron}
\bibinfo{author}{{Baron}, D.}, \bibinfo{year}{2019}.
\newblock \bibinfo{title}{{Machine Learning in Astronomy: a practical
  overview}}.
\newblock \bibinfo{journal}{arXiv e-prints} ,
  \bibinfo{pages}{arXiv:1904.07248}\href{http://arxiv.org/abs/1904.07248}{\tt
  arXiv:1904.07248}.
%Type = Article
\bibitem[{{Baron} et~al.(2015){Baron}, {Poznanski}, {Watson}, {Yao}, {Cox} and
  {Prochaska}}]{2015Baron}
\bibinfo{author}{{Baron}, D.}, \bibinfo{author}{{Poznanski}, D.},
  \bibinfo{author}{{Watson}, D.}, \bibinfo{author}{{Yao}, Y.},
  \bibinfo{author}{{Cox}, N.L.J.}, \bibinfo{author}{{Prochaska}, J.X.},
  \bibinfo{year}{2015}.
\newblock \bibinfo{title}{{Using Machine Learning to classify the diffuse
  interstellar bands}}.
\newblock \bibinfo{journal}{Monthly Notices of the Royal Astronomy Society}
  \bibinfo{volume}{451}, \bibinfo{pages}{332--352}.
\newblock \DOIprefix\doi{10.1093/mnras/stv977},
  \href{http://arxiv.org/abs/1501.04631}{\tt arXiv:1501.04631}.
%Type = Article
\bibitem[{{Carruba} et~al.(2016){Carruba}, {Aljbaae} and
  {Winter}}]{2016Carruba}
\bibinfo{author}{{Carruba}, V.}, \bibinfo{author}{{Aljbaae}, S.},
  \bibinfo{author}{{Winter}, O.C.}, \bibinfo{year}{2016}.
\newblock \bibinfo{title}{{On the Erigone family and the z$_{2}$ secular
  resonance}}.
\newblock \bibinfo{journal}{Monthly Notices of the Royal Astronomy Society}
  \bibinfo{volume}{455}, \bibinfo{pages}{2279--2288}.
\newblock \DOIprefix\doi{10.1093/mnras/stv2430},
  \href{http://arxiv.org/abs/1510.05551}{\tt arXiv:1510.05551}.
%Type = Article
\bibitem[{{Carruba} and {Michtchenko}(2007)}]{2007Carruba}
\bibinfo{author}{{Carruba}, V.}, \bibinfo{author}{{Michtchenko}, T.A.},
  \bibinfo{year}{2007}.
\newblock \bibinfo{title}{{A frequency approach to identifying asteroid
  families}}.
\newblock \bibinfo{journal}{Astronomy and Astrophysics} \bibinfo{volume}{475},
  \bibinfo{pages}{1145--1158}.
\newblock \DOIprefix\doi{10.1051/0004-6361:20077689}.
%Type = Article
\bibitem[{{Carruba} and {Michtchenko}(2009)}]{2009Carruba}
\bibinfo{author}{{Carruba}, V.}, \bibinfo{author}{{Michtchenko}, T.A.},
  \bibinfo{year}{2009}.
\newblock \bibinfo{title}{{A frequency approach to identifying asteroid
  families. II. Families interacting with nonlinear secular resonances and
  low-order mean-motion resonances}}.
\newblock \bibinfo{journal}{Astronomy and Astrophysics} \bibinfo{volume}{493},
  \bibinfo{pages}{267--282}.
\newblock \DOIprefix\doi{10.1051/0004-6361:200809852}.
%Type = Article
\bibitem[{{Carruba} et~al.(2015){Carruba}, {Nesvorn{\'y}}, {Aljbaae} and
  {Huaman}}]{2015Carruba}
\bibinfo{author}{{Carruba}, V.}, \bibinfo{author}{{Nesvorn{\'y}}, D.},
  \bibinfo{author}{{Aljbaae}, S.}, \bibinfo{author}{{Huaman}, M.E.},
  \bibinfo{year}{2015}.
\newblock \bibinfo{title}{{Dynamical evolution of the Cybele asteroids}}.
\newblock \bibinfo{journal}{Monthly Notices of the Royal Astronomy Society}
  \bibinfo{volume}{451}, \bibinfo{pages}{244--256}.
\newblock \DOIprefix\doi{10.1093/mnras/stv997},
  \href{http://arxiv.org/abs/1505.03745}{\tt arXiv:1505.03745}.
%Type = Article
\bibitem[{{Castro-Ginard} et~al.(2018){Castro-Ginard}, {Jordi}, {Luri},
  {Julbe}, {Morvan}, {Balaguer-N{\'u}{\~n}ez} and
  {Cantat-Gaudin}}]{2018A&A...618A..59C}
\bibinfo{author}{{Castro-Ginard}, A.}, \bibinfo{author}{{Jordi}, C.},
  \bibinfo{author}{{Luri}, X.}, \bibinfo{author}{{Julbe}, F.},
  \bibinfo{author}{{Morvan}, M.}, \bibinfo{author}{{Balaguer-N{\'u}{\~n}ez},
  L.}, \bibinfo{author}{{Cantat-Gaudin}, T.}, \bibinfo{year}{2018}.
\newblock \bibinfo{title}{{A new method for unveiling open clusters in Gaia.
  New nearby open clusters confirmed by DR2}}.
\newblock \bibinfo{journal}{Astronomy and Astrophysics} \bibinfo{volume}{618},
  \bibinfo{pages}{A59}.
\newblock \DOIprefix\doi{10.1051/0004-6361/201833390},
  \href{http://arxiv.org/abs/1805.03045}{\tt arXiv:1805.03045}.
%Type = Article
\bibitem[{{Chen} et~al.(2020){Chen}, {Li}, {Yuan}, {Huang}, {Tian}, {Wang},
  {Zhang}, {Wang} and {Liu}}]{2020MNRAS.493..351C}
\bibinfo{author}{{Chen}, B.Q.}, \bibinfo{author}{{Li}, G.X.},
  \bibinfo{author}{{Yuan}, H.B.}, \bibinfo{author}{{Huang}, Y.},
  \bibinfo{author}{{Tian}, Z.J.}, \bibinfo{author}{{Wang}, H.F.},
  \bibinfo{author}{{Zhang}, H.W.}, \bibinfo{author}{{Wang}, C.},
  \bibinfo{author}{{Liu}, X.W.}, \bibinfo{year}{2020}.
\newblock \bibinfo{title}{{A large catalogue of molecular clouds with accurate
  distances within 4 kpc of the Galactic disc}}.
\newblock \bibinfo{journal}{Monthly Notices of the Royal Astronomy Society}
  \bibinfo{volume}{493}, \bibinfo{pages}{351--361}.
\newblock \DOIprefix\doi{10.1093/mnras/staa235},
  \href{http://arxiv.org/abs/2001.11682}{\tt arXiv:2001.11682}.
%Type = Article
\bibitem[{{Chon} et~al.(2013){Chon}, {B{\"o}hringer} and {Nowak}}]{2013Chon}
\bibinfo{author}{{Chon}, G.}, \bibinfo{author}{{B{\"o}hringer}, H.},
  \bibinfo{author}{{Nowak}, N.}, \bibinfo{year}{2013}.
\newblock \bibinfo{title}{{The extended ROSAT-ESO Flux-Limited X-ray Galaxy
  Cluster Survey (REFLEX II) - III. Construction of the first flux-limited
  supercluster sample}}.
\newblock \bibinfo{journal}{Monthly Notices of the Royal Astronomy Society}
  \bibinfo{volume}{429}, \bibinfo{pages}{3272--3287}.
\newblock \DOIprefix\doi{10.1093/mnras/sts584},
  \href{http://arxiv.org/abs/1212.1597}{\tt arXiv:1212.1597}.
%Type = Article
\bibitem[{{Chow-Mart{\'\i}nez} et~al.(2014){Chow-Mart{\'\i}nez}, {Andernach},
  {Caretta} and {Trejo-Alonso}}]{2014Chow}
\bibinfo{author}{{Chow-Mart{\'\i}nez}, M.}, \bibinfo{author}{{Andernach}, H.},
  \bibinfo{author}{{Caretta}, C.A.}, \bibinfo{author}{{Trejo-Alonso}, J.J.},
  \bibinfo{year}{2014}.
\newblock \bibinfo{title}{{Two new catalogues of superclusters of Abell/ACO
  galaxy clusters out to redshift 0.15}}.
\newblock \bibinfo{journal}{Monthly Notices of the Royal Astronomy Society}
  \bibinfo{volume}{445}, \bibinfo{pages}{4073--4085}.
\newblock \DOIprefix\doi{10.1093/mnras/stu1961},
  \href{http://arxiv.org/abs/1409.5152}{\tt arXiv:1409.5152}.
%Type = Article
\bibitem[{{Colombo} et~al.(2015){Colombo}, {Rosolowsky}, {Ginsburg},
  {Duarte-Cabral} and {Hughes}}]{2015MNRAS.454.2067C}
\bibinfo{author}{{Colombo}, D.}, \bibinfo{author}{{Rosolowsky}, E.},
  \bibinfo{author}{{Ginsburg}, A.}, \bibinfo{author}{{Duarte-Cabral}, A.},
  \bibinfo{author}{{Hughes}, A.}, \bibinfo{year}{2015}.
\newblock \bibinfo{title}{{Graph-based interpretation of the molecular
  interstellar medium segmentation}}.
\newblock \bibinfo{journal}{Monthly Notices of the Royal Astronomy Society}
  \bibinfo{volume}{454}, \bibinfo{pages}{2067--2091}.
\newblock \DOIprefix\doi{10.1093/mnras/stv2063},
  \href{http://arxiv.org/abs/1510.04253}{\tt arXiv:1510.04253}.
%Type = Article
\bibitem[{{de Souza} and {Ciardi}(2015)}]{2015Souza}
\bibinfo{author}{{de Souza}, R.S.}, \bibinfo{author}{{Ciardi}, B.},
  \bibinfo{year}{2015}.
\newblock \bibinfo{title}{{AMADA-Analysis of multidimensional astronomical
  datasets}}.
\newblock \bibinfo{journal}{Astronomy and Computing} \bibinfo{volume}{12},
  \bibinfo{pages}{100--108}.
\newblock \DOIprefix\doi{10.1016/j.ascom.2015.06.006},
  \href{http://arxiv.org/abs/1503.07736}{\tt arXiv:1503.07736}.
%Type = Article
\bibitem[{Defays(1977)}]{Defays77}
\bibinfo{author}{Defays, D.}, \bibinfo{year}{1977}.
\newblock \bibinfo{title}{An efficient algorithm for a complete link method}.
\newblock \bibinfo{journal}{The Computer Journal} \bibinfo{volume}{20},
  \bibinfo{pages}{364--366}.
%Type = Article
\bibitem[{Diaferio(1999)}]{Diaferio1999}
\bibinfo{author}{Diaferio, A.}, \bibinfo{year}{1999}.
\newblock \bibinfo{title}{Mass estimation in the outer regions of galaxy
  clusters}.
\newblock \bibinfo{journal}{Monthly Notices of the Royal Astronomical Society}
  \bibinfo{volume}{309}, \bibinfo{pages}{610--622}.
\newblock \URLprefix \url{http://adsabs.harvard.edu/abs/1999MNRAS.309..610D}.
%Type = Article
\bibitem[{{Dias} et~al.(2006){Dias}, {Assafin}, {Fl{\'o}rio}, {Alessi} and
  {L{\'\i}bero}}]{2006A&A...446..949D}
\bibinfo{author}{{Dias}, W.S.}, \bibinfo{author}{{Assafin}, M.},
  \bibinfo{author}{{Fl{\'o}rio}, V.}, \bibinfo{author}{{Alessi}, B.S.},
  \bibinfo{author}{{L{\'\i}bero}, V.}, \bibinfo{year}{2006}.
\newblock \bibinfo{title}{{Proper motion determination of open clusters based
  on the UCAC2 catalogue}}.
\newblock \bibinfo{journal}{Astronomy and Astrophysics} \bibinfo{volume}{446},
  \bibinfo{pages}{949--953}.
\newblock \DOIprefix\doi{10.1051/0004-6361:20052741}.
%Type = Article
\bibitem[{{Eigenson} and {Yatsyk}(1987)}]{1987Eigenson}
\bibinfo{author}{{Eigenson}, A.M.}, \bibinfo{author}{{Yatsyk}, O.S.},
  \bibinfo{year}{1987}.
\newblock \bibinfo{title}{{Star Cluster Taxonomy}}.
\newblock \bibinfo{journal}{Soviet Astronomy Letters} \bibinfo{volume}{13},
  \bibinfo{pages}{197}.
%Type = Article
\bibitem[{{Einasto} et~al.(2007){Einasto}, {Einasto}, {Tago}, {Saar},
  {H{\"u}tsi}, {J{\~o}eveer}, {Liivam{\"a}gi}, {Suhhonenko}, {Jaaniste},
  {Hein{\"a}m{\"a}ki}, {M{\"u}ller}, {Knebe} and
  {Tucker}}]{2007A&A...462..811E}
\bibinfo{author}{{Einasto}, J.}, \bibinfo{author}{{Einasto}, M.},
  \bibinfo{author}{{Tago}, E.}, \bibinfo{author}{{Saar}, E.},
  \bibinfo{author}{{H{\"u}tsi}, G.}, \bibinfo{author}{{J{\~o}eveer}, M.},
  \bibinfo{author}{{Liivam{\"a}gi}, L.J.}, \bibinfo{author}{{Suhhonenko}, I.},
  \bibinfo{author}{{Jaaniste}, J.}, \bibinfo{author}{{Hein{\"a}m{\"a}ki}, P.},
  \bibinfo{author}{{M{\"u}ller}, V.}, \bibinfo{author}{{Knebe}, A.},
  \bibinfo{author}{{Tucker}, D.}, \bibinfo{year}{2007}.
\newblock \bibinfo{title}{{Superclusters of galaxies from the 2dF redshift
  survey. I. The catalogue}}.
\newblock \bibinfo{journal}{Astronomy and Astrophysics} \bibinfo{volume}{462},
  \bibinfo{pages}{811--825}.
\newblock \DOIprefix\doi{10.1051/0004-6361:20065296},
  \href{http://arxiv.org/abs/astro-ph/0603764}{\tt arXiv:astro-ph/0603764}.
%Type = Book
\bibitem[{Everitt et~al.(2011)Everitt, Landau, Leese and Stahl}]{Everitt2011}
\bibinfo{author}{Everitt, B.S.}, \bibinfo{author}{Landau, S.},
  \bibinfo{author}{Leese, M.}, \bibinfo{author}{Stahl, D.},
  \bibinfo{year}{2011}.
\newblock \bibinfo{title}{Cluster Analysis, 5th Edition}.
\newblock \bibinfo{publisher}{Wiley Online Library}, \bibinfo{address}{Hoboken,
  NJ}.
%Type = Article
\bibitem[{{Galli} et~al.(2019){Galli}, {Loinard}, {Bouy}, {Sarro},
  {Ortiz-Le{\'o}n}, {Dzib}, {Olivares}, {Heyer}, {Hernandez},
  {Rom{\'a}n-Z{\'u}{\~n}iga}, {Kounkel} and {Covey}}]{2019A&A...630A.137G}
\bibinfo{author}{{Galli}, P.A.B.}, \bibinfo{author}{{Loinard}, L.},
  \bibinfo{author}{{Bouy}, H.}, \bibinfo{author}{{Sarro}, L.M.},
  \bibinfo{author}{{Ortiz-Le{\'o}n}, G.N.}, \bibinfo{author}{{Dzib}, S.A.},
  \bibinfo{author}{{Olivares}, J.}, \bibinfo{author}{{Heyer}, M.},
  \bibinfo{author}{{Hernandez}, J.},
  \bibinfo{author}{{Rom{\'a}n-Z{\'u}{\~n}iga}, C.}, \bibinfo{author}{{Kounkel},
  M.}, \bibinfo{author}{{Covey}, K.}, \bibinfo{year}{2019}.
\newblock \bibinfo{title}{{Structure and kinematics of the Taurus star-forming
  region from Gaia-DR2 and VLBI astrometry}}.
\newblock \bibinfo{journal}{Astronomy and Astrophysics} \bibinfo{volume}{630},
  \bibinfo{pages}{A137}.
\newblock \DOIprefix\doi{10.1051/0004-6361/201935928},
  \href{http://arxiv.org/abs/1909.01118}{\tt arXiv:1909.01118}.
%Type = Article
\bibitem[{{Gao}(2014)}]{2014RAA....14..159G}
\bibinfo{author}{{Gao}, X.H.}, \bibinfo{year}{2014}.
\newblock \bibinfo{title}{{Membership determination of open cluster NGC 188
  based on the DBSCAN clustering algorithm}}.
\newblock \bibinfo{journal}{Research in Astronomy and Astrophysics}
  \bibinfo{volume}{14}, \bibinfo{pages}{159--164}.
\newblock \DOIprefix\doi{10.1088/1674-4527/14/2/004}.
%Type = Article
\bibitem[{{Garcia}(1993)}]{1993Garcia}
\bibinfo{author}{{Garcia}, A.M.}, \bibinfo{year}{1993}.
\newblock \bibinfo{title}{{General study of group membership. II. Determination
  of nearby groups.}}
\newblock \bibinfo{journal}{Astronomy and Astrophysicss} \bibinfo{volume}{100},
  \bibinfo{pages}{47--90}.
%Type = Article
\bibitem[{{Garcia} et~al.(1992){Garcia}, {Morenas} and {Paturel}}]{1992Garcia}
\bibinfo{author}{{Garcia}, A.M.}, \bibinfo{author}{{Morenas}, V.},
  \bibinfo{author}{{Paturel}, G.}, \bibinfo{year}{1992}.
\newblock \bibinfo{title}{{A search for a quantitative comparison of galaxy
  clustering algorithms}}.
\newblock \bibinfo{journal}{Astronomy and Astrophysics} \bibinfo{volume}{253},
  \bibinfo{pages}{74--76}.
%Type = Article
\bibitem[{{Goodman} et~al.(2009){Goodman}, {Rosolowsky}, {Borkin}, {Foster},
  {Halle}, {Kauffmann} and {Pineda}}]{2009Goodman}
\bibinfo{author}{{Goodman}, A.A.}, \bibinfo{author}{{Rosolowsky}, E.W.},
  \bibinfo{author}{{Borkin}, M.A.}, \bibinfo{author}{{Foster}, J.B.},
  \bibinfo{author}{{Halle}, M.}, \bibinfo{author}{{Kauffmann}, J.},
  \bibinfo{author}{{Pineda}, J.E.}, \bibinfo{year}{2009}.
\newblock \bibinfo{title}{{A role for self-gravity at multiple length scales in
  the process of star formation}}.
\newblock \bibinfo{journal}{Nature} \bibinfo{volume}{457},
  \bibinfo{pages}{63--66}.
\newblock \DOIprefix\doi{10.1038/nature07609}.
%Type = Article
\bibitem[{Gordon(1987)}]{Gordon1987}
\bibinfo{author}{Gordon, A.D.}, \bibinfo{year}{1987}.
\newblock \bibinfo{title}{A review of hierarchical classification}.
\newblock \bibinfo{journal}{Journal of the Royal Statistical Society: Series A
  (General)} \bibinfo{volume}{150}, \bibinfo{pages}{119--137}.
\newblock \URLprefix
  \url{https://rss.onlinelibrary.wiley.com/doi/abs/10.2307/2981629},
  \DOIprefix\doi{https://doi.org/10.2307/2981629},
  \href{http://arxiv.org/abs/https://rss.onlinelibrary.wiley.com/doi/pdf/10.2307/2981629}{\tt
  arXiv:https://rss.onlinelibrary.wiley.com/doi/pdf/10.2307/2981629}.
%Type = Article
\bibitem[{Gourgoulhon et~al.(1992)Gourgoulhon, Chamaraux and
  Fouque}]{Gourgoulhon1992}
\bibinfo{author}{Gourgoulhon, E.}, \bibinfo{author}{Chamaraux, P.},
  \bibinfo{author}{Fouque, P.}, \bibinfo{year}{1992}.
\newblock \bibinfo{title}{Groups of galaxies within 80 {Mpc}. {I} - {Grouping}
  hierarchical method and statistical properties}.
\newblock \bibinfo{journal}{Astronomy and Astrophysics} \bibinfo{volume}{255},
  \bibinfo{pages}{69--86}.
%Type = Article
\bibitem[{{Guennou} et~al.(2014){Guennou}, {Adami}, {Durret}, {Lima Neto},
  {Ulmer}, {Clowe}, {LeBrun}, {Martinet}, {Allam}, {Annis}, {Basa}, {Benoist},
  {Biviano}, {Cappi}, {Cypriano}, {Gavazzi}, {Halliday}, {Ilbert}, {Jullo},
  {Just}, {Limousin}, {M{\'a}rquez}, {Mazure}, {Murphy}, {Plana}, {Rostagni},
  {Russeil}, {Schirmer}, {Slezak}, {Tucker}, {Zaritsky} and
  {Ziegler}}]{2014A&A...561A.112G}
\bibinfo{author}{{Guennou}, L.}, \bibinfo{author}{{Adami}, C.},
  \bibinfo{author}{{Durret}, F.}, \bibinfo{author}{{Lima Neto}, G.B.},
  \bibinfo{author}{{Ulmer}, M.P.}, \bibinfo{author}{{Clowe}, D.},
  \bibinfo{author}{{LeBrun}, V.}, \bibinfo{author}{{Martinet}, N.},
  \bibinfo{author}{{Allam}, S.}, \bibinfo{author}{{Annis}, J.},
  \bibinfo{author}{{Basa}, S.}, \bibinfo{author}{{Benoist}, C.},
  \bibinfo{author}{{Biviano}, A.}, \bibinfo{author}{{Cappi}, A.},
  \bibinfo{author}{{Cypriano}, E.S.}, \bibinfo{author}{{Gavazzi}, R.},
  \bibinfo{author}{{Halliday}, C.}, \bibinfo{author}{{Ilbert}, O.},
  \bibinfo{author}{{Jullo}, E.}, \bibinfo{author}{{Just}, D.},
  \bibinfo{author}{{Limousin}, M.}, \bibinfo{author}{{M{\'a}rquez}, I.},
  \bibinfo{author}{{Mazure}, A.}, \bibinfo{author}{{Murphy}, K.J.},
  \bibinfo{author}{{Plana}, H.}, \bibinfo{author}{{Rostagni}, F.},
  \bibinfo{author}{{Russeil}, D.}, \bibinfo{author}{{Schirmer}, M.},
  \bibinfo{author}{{Slezak}, E.}, \bibinfo{author}{{Tucker}, D.},
  \bibinfo{author}{{Zaritsky}, D.}, \bibinfo{author}{{Ziegler}, B.},
  \bibinfo{year}{2014}.
\newblock \bibinfo{title}{{Structure and substructure analysis of DAFT/FADA
  galaxy clusters in the [0.4-0.9] redshift range}}.
\newblock \bibinfo{journal}{Astronomy and Astrophysics} \bibinfo{volume}{561},
  \bibinfo{pages}{A112}.
\newblock \DOIprefix\doi{10.1051/0004-6361/201321208},
  \href{http://arxiv.org/abs/1311.6922}{\tt arXiv:1311.6922}.
%Type = Article
\bibitem[{{Guo} et~al.(2022){Guo}, {Chen} and {Liu}}]{2022MNRAS.511.2302G}
\bibinfo{author}{{Guo}, H.L.}, \bibinfo{author}{{Chen}, B.Q.},
  \bibinfo{author}{{Liu}, X.W.}, \bibinfo{year}{2022}.
\newblock \bibinfo{title}{{A large catalogue of molecular clouds in the
  Southern sky}}.
\newblock \bibinfo{journal}{Monthly Notices of the Royal Astronomy Society}
  \bibinfo{volume}{511}, \bibinfo{pages}{2302--2312}.
\newblock \DOIprefix\doi{10.1093/mnras/stac213},
  \href{http://arxiv.org/abs/2201.09578}{\tt arXiv:2201.09578}.
%Type = Article
\bibitem[{{Henshaw} et~al.(2019){Henshaw}, {Ginsburg}, {Haworth}, {Longmore},
  {Kruijssen}, {Mills}, {Sokolov}, {Walker}, {Barnes}, {Contreras}, {Bally},
  {Battersby}, {Beuther}, {Butterfield}, {Dale}, {Henning}, {Jackson},
  {Kauffmann}, {Pillai}, {Ragan}, {Riener} and {Zhang}}]{2019MNRAS.485.2457H}
\bibinfo{author}{{Henshaw}, J.D.}, \bibinfo{author}{{Ginsburg}, A.},
  \bibinfo{author}{{Haworth}, T.J.}, \bibinfo{author}{{Longmore}, S.N.},
  \bibinfo{author}{{Kruijssen}, J.M.D.}, \bibinfo{author}{{Mills}, E.A.C.},
  \bibinfo{author}{{Sokolov}, V.}, \bibinfo{author}{{Walker}, D.L.},
  \bibinfo{author}{{Barnes}, A.T.}, \bibinfo{author}{{Contreras}, Y.},
  \bibinfo{author}{{Bally}, J.}, \bibinfo{author}{{Battersby}, C.},
  \bibinfo{author}{{Beuther}, H.}, \bibinfo{author}{{Butterfield}, N.},
  \bibinfo{author}{{Dale}, J.E.}, \bibinfo{author}{{Henning}, T.},
  \bibinfo{author}{{Jackson}, J.M.}, \bibinfo{author}{{Kauffmann}, J.},
  \bibinfo{author}{{Pillai}, T.}, \bibinfo{author}{{Ragan}, S.},
  \bibinfo{author}{{Riener}, M.}, \bibinfo{author}{{Zhang}, Q.},
  \bibinfo{year}{2019}.
\newblock \bibinfo{title}{{`The Brick' is not a brick: a comprehensive study of
  the structure and dynamics of the central molecular zone cloud
  G0.253+0.016}}.
\newblock \bibinfo{journal}{Monthly Notices of the Royal Astronomy Society}
  \bibinfo{volume}{485}, \bibinfo{pages}{2457--2485}.
\newblock \DOIprefix\doi{10.1093/mnras/stz471},
  \href{http://arxiv.org/abs/1902.02793}{\tt arXiv:1902.02793}.
%Type = Article
\bibitem[{{Hocking} et~al.(2018){Hocking}, {Geach}, {Sun} and
  {Davey}}]{2018Hocking}
\bibinfo{author}{{Hocking}, A.}, \bibinfo{author}{{Geach}, J.E.},
  \bibinfo{author}{{Sun}, Y.}, \bibinfo{author}{{Davey}, N.},
  \bibinfo{year}{2018}.
\newblock \bibinfo{title}{{An automatic taxonomy of galaxy morphology using
  unsupervised machine learning}}.
\newblock \bibinfo{journal}{Monthly Notices of the Royal Astronomy Society}
  \bibinfo{volume}{473}, \bibinfo{pages}{1108--1129}.
\newblock \DOIprefix\doi{10.1093/mnras/stx2351},
  \href{http://arxiv.org/abs/1709.05834}{\tt arXiv:1709.05834}.
%Type = Article
\bibitem[{{Hojnacki} et~al.(2007){Hojnacki}, {Kastner}, {Micela}, {Feigelson}
  and {LaLonde}}]{2007Hojnacki}
\bibinfo{author}{{Hojnacki}, S.M.}, \bibinfo{author}{{Kastner}, J.H.},
  \bibinfo{author}{{Micela}, G.}, \bibinfo{author}{{Feigelson}, E.D.},
  \bibinfo{author}{{LaLonde}, S.M.}, \bibinfo{year}{2007}.
\newblock \bibinfo{title}{{An X-Ray Spectral Classification Algorithm with
  Application to Young Stellar Clusters}}.
\newblock \bibinfo{journal}{Astrophysics Journal} \bibinfo{volume}{659},
  \bibinfo{pages}{585--598}.
\newblock \DOIprefix\doi{10.1086/512232}.
%Type = Article
\bibitem[{{Houlahan} and {Scalo}(1992)}]{1992ApJ...393..172H}
\bibinfo{author}{{Houlahan}, P.}, \bibinfo{author}{{Scalo}, J.},
  \bibinfo{year}{1992}.
\newblock \bibinfo{title}{{Recognition and Characterization of Hierarchical
  Interstellar Structure. II. Structure Tree Statistics}}.
\newblock \bibinfo{journal}{Astrophysics Journal} \bibinfo{volume}{393},
  \bibinfo{pages}{172}.
\newblock \DOIprefix\doi{10.1086/171495}.
%Type = Article
\bibitem[{{Hsu} et~al.(2013){Hsu}, {Ebeling} and {Richard}}]{MACS0358}
\bibinfo{author}{{Hsu}, L.Y.}, \bibinfo{author}{{Ebeling}, H.},
  \bibinfo{author}{{Richard}, J.}, \bibinfo{year}{2013}.
\newblock \bibinfo{title}{{The three-dimensional geometry and merger history of
  the massive galaxy cluster MACS J0358.8-2955}}.
\newblock \bibinfo{journal}{Monthly Notices of the Royal Astronomy Society}
  \bibinfo{volume}{429}, \bibinfo{pages}{833--848}.
\newblock \DOIprefix\doi{10.1093/mnras/sts379},
  \href{http://arxiv.org/abs/1209.2492}{\tt arXiv:1209.2492}.
%Type = Article
\bibitem[{{Huchra} and {Geller}(1982)}]{1982Huchra}
\bibinfo{author}{{Huchra}, J.P.}, \bibinfo{author}{{Geller}, M.J.},
  \bibinfo{year}{1982}.
\newblock \bibinfo{title}{{Groups of Galaxies. I. Nearby groups}}.
\newblock \bibinfo{journal}{Astrophysics Journal} \bibinfo{volume}{257},
  \bibinfo{pages}{423--437}.
\newblock \DOIprefix\doi{10.1086/160000}.
%Type = Book
\bibitem[{{Ivezi{\'c}} et~al.(2014){Ivezi{\'c}}, {Connolly}, {VanderPlas} and
  {Gray}}]{Ivezic2014}
\bibinfo{author}{{Ivezi{\'c}}, {\v{Z}}.}, \bibinfo{author}{{Connolly}, A.J.},
  \bibinfo{author}{{VanderPlas}, J.T.}, \bibinfo{author}{{Gray}, A.},
  \bibinfo{year}{2014}.
\newblock \bibinfo{title}{{Statistics, Data Mining, and Machine Learning in
  Astronomy: A Practical Python Guide for the Analysis of Survey Data}}.
\newblock \bibinfo{publisher}{Princeton University Press},
  \bibinfo{address}{Princeton, NJ}.
\newblock \DOIprefix\doi{10.1515/9781400848911}.
%Type = Inproceedings
\bibitem[{{Kne{\v{z}}evi{\'c}}(2016)}]{2016IAUS..318...16K}
\bibinfo{author}{{Kne{\v{z}}evi{\'c}}, Z.}, \bibinfo{year}{2016}.
\newblock \bibinfo{title}{{Asteroid Family Identification: History and State of
  the Art}}, in: \bibinfo{editor}{{Chesley}, S.R.},
  \bibinfo{editor}{{Morbidelli}, A.}, \bibinfo{editor}{{Jedicke}, R.},
  \bibinfo{editor}{{Farnocchia}, D.} (Eds.), \bibinfo{booktitle}{Asteroids: New
  Observations, New Models}, pp. \bibinfo{pages}{16--27}.
\newblock \DOIprefix\doi{10.1017/S1743921315008728}.
%Type = Article
\bibitem[{{Krone-Martins} and {Moitinho}(2014)}]{2014A&A...561A..57K}
\bibinfo{author}{{Krone-Martins}, A.}, \bibinfo{author}{{Moitinho}, A.},
  \bibinfo{year}{2014}.
\newblock \bibinfo{title}{{UPMASK: unsupervised photometric membership
  assignment in stellar clusters}}.
\newblock \bibinfo{journal}{Astronomy and Astrophysics} \bibinfo{volume}{561},
  \bibinfo{pages}{A57}.
\newblock \DOIprefix\doi{10.1051/0004-6361/201321143},
  \href{http://arxiv.org/abs/1309.4471}{\tt arXiv:1309.4471}.
%Type = Article
\bibitem[{{Krone-Martins} et~al.(2010){Krone-Martins}, {Soubiran}, {Ducourant},
  {Teixeira} and {Le Campion}}]{2010A&A...516A...3K}
\bibinfo{author}{{Krone-Martins}, A.}, \bibinfo{author}{{Soubiran}, C.},
  \bibinfo{author}{{Ducourant}, C.}, \bibinfo{author}{{Teixeira}, R.},
  \bibinfo{author}{{Le Campion}, J.F.}, \bibinfo{year}{2010}.
\newblock \bibinfo{title}{{Kinematic parameters and membership probabilities of
  open clusters in the Bordeaux PM2000 catalogue}}.
\newblock \bibinfo{journal}{Astronomy and Astrophysics} \bibinfo{volume}{516},
  \bibinfo{pages}{A3}.
\newblock \DOIprefix\doi{10.1051/0004-6361/200913881},
  \href{http://arxiv.org/abs/1006.0096}{\tt arXiv:1006.0096}.
%Type = Article
\bibitem[{Lance and Williams(1967)}]{Lance1967}
\bibinfo{author}{Lance, G.N.}, \bibinfo{author}{Williams, W.T.},
  \bibinfo{year}{1967}.
\newblock \bibinfo{title}{{A General Theory of Classificatory Sorting
  Strategies: 1. Hierarchical Systems}}.
\newblock \bibinfo{journal}{The Computer Journal} \bibinfo{volume}{9},
  \bibinfo{pages}{373--380}.
\newblock \URLprefix \url{https://doi.org/10.1093/comjnl/9.4.373},
  \DOIprefix\doi{10.1093/comjnl/9.4.373},
  \href{http://arxiv.org/abs/https://academic.oup.com/comjnl/article-pdf/9/4/373/1101470/9-4-373.pdf}{\tt
  arXiv:https://academic.oup.com/comjnl/article-pdf/9/4/373/1101470/9-4-373.pdf}.
%Type = Article
\bibitem[{{Liivam{\"a}gi} et~al.(2012){Liivam{\"a}gi}, {Tempel} and
  {Saar}}]{2012A&A...539A..80L}
\bibinfo{author}{{Liivam{\"a}gi}, L.J.}, \bibinfo{author}{{Tempel}, E.},
  \bibinfo{author}{{Saar}, E.}, \bibinfo{year}{2012}.
\newblock \bibinfo{title}{{SDSS DR7 superclusters. The catalogues}}.
\newblock \bibinfo{journal}{Astronomy and Astrophysics} \bibinfo{volume}{539},
  \bibinfo{pages}{A80}.
\newblock \DOIprefix\doi{10.1051/0004-6361/201016288},
  \href{http://arxiv.org/abs/1012.1989}{\tt arXiv:1012.1989}.
%Type = Article
\bibitem[{{Liu} et~al.(2021){Liu}, {Bulbul}, {Ghirardini}, {Liu}, {Klein},
  {Clerc}, {Oezsoy}, {Ramos-Ceja}, {Pacaud}, {Comparat}, {Okabe}, {Bahar},
  {Biffi}, {Brunner}, {Brueggen}, {Buchner}, {Ider Chitham}, {Chiu}, {Dolag},
  {Gatuzz}, {Gonzalez}, {Hoang}, {Lamer}, {Merloni}, {Nandra}, {Oguri}, {Ota},
  {Predehl}, {Reiprich}, {Salvato}, {Schrabback}, {Sanders}, {Seppi} and
  {Thibaud}}]{2021Liu}
\bibinfo{author}{{Liu}, A.}, \bibinfo{author}{{Bulbul}, E.},
  \bibinfo{author}{{Ghirardini}, V.}, \bibinfo{author}{{Liu}, T.},
  \bibinfo{author}{{Klein}, M.}, \bibinfo{author}{{Clerc}, N.},
  \bibinfo{author}{{Oezsoy}, Y.}, \bibinfo{author}{{Ramos-Ceja}, M.E.},
  \bibinfo{author}{{Pacaud}, F.}, \bibinfo{author}{{Comparat}, J.},
  \bibinfo{author}{{Okabe}, N.}, \bibinfo{author}{{Bahar}, Y.E.},
  \bibinfo{author}{{Biffi}, V.}, \bibinfo{author}{{Brunner}, H.},
  \bibinfo{author}{{Brueggen}, M.}, \bibinfo{author}{{Buchner}, J.},
  \bibinfo{author}{{Ider Chitham}, J.}, \bibinfo{author}{{Chiu}, I.},
  \bibinfo{author}{{Dolag}, K.}, \bibinfo{author}{{Gatuzz}, E.},
  \bibinfo{author}{{Gonzalez}, J.}, \bibinfo{author}{{Hoang}, D.N.},
  \bibinfo{author}{{Lamer}, G.}, \bibinfo{author}{{Merloni}, A.},
  \bibinfo{author}{{Nandra}, K.}, \bibinfo{author}{{Oguri}, M.},
  \bibinfo{author}{{Ota}, N.}, \bibinfo{author}{{Predehl}, P.},
  \bibinfo{author}{{Reiprich}, T.H.}, \bibinfo{author}{{Salvato}, M.},
  \bibinfo{author}{{Schrabback}, T.}, \bibinfo{author}{{Sanders}, J.S.},
  \bibinfo{author}{{Seppi}, R.}, \bibinfo{author}{{Thibaud}, Q.},
  \bibinfo{year}{2021}.
\newblock \bibinfo{title}{{The eROSITA Final Equatorial-Depth Survey (eFEDS):
  Catalog of galaxy clusters and groups}}.
\newblock \bibinfo{journal}{arXiv e-prints} ,
  \bibinfo{pages}{arXiv:2106.14518}\href{http://arxiv.org/abs/2106.14518}{\tt
  arXiv:2106.14518}.
%Type = Article
\bibitem[{{Liu} et~al.(2018){Liu}, {Yu}, {Diaferio}, {Tozzi}, {Hwang},
  {Umetsu}, {Okabe} and {Yang}}]{2018ApJ...863..102L}
\bibinfo{author}{{Liu}, A.}, \bibinfo{author}{{Yu}, H.},
  \bibinfo{author}{{Diaferio}, A.}, \bibinfo{author}{{Tozzi}, P.},
  \bibinfo{author}{{Hwang}, H.S.}, \bibinfo{author}{{Umetsu}, K.},
  \bibinfo{author}{{Okabe}, N.}, \bibinfo{author}{{Yang}, L.L.},
  \bibinfo{year}{2018}.
\newblock \bibinfo{title}{{Inside a Beehive: The Multiple Merging Processes in
  the Galaxy Cluster Abell 2142}}.
\newblock \bibinfo{journal}{Astrophysics Journal} \bibinfo{volume}{863},
  \bibinfo{pages}{102}.
\newblock \DOIprefix\doi{10.3847/1538-4357/aad090},
  \href{http://arxiv.org/abs/1806.10864}{\tt arXiv:1806.10864}.
%Type = Article
\bibitem[{{Luparello} et~al.(2011){Luparello}, {Lares}, {Lambas} and
  {Padilla}}]{2011MNRAS.415..964L}
\bibinfo{author}{{Luparello}, H.}, \bibinfo{author}{{Lares}, M.},
  \bibinfo{author}{{Lambas}, D.G.}, \bibinfo{author}{{Padilla}, N.},
  \bibinfo{year}{2011}.
\newblock \bibinfo{title}{{Future virialized structures: an analysis of
  superstructures in the SDSS-DR7}}.
\newblock \bibinfo{journal}{Monthly Notices of the Royal Astronomy Society}
  \bibinfo{volume}{415}, \bibinfo{pages}{964--976}.
\newblock \DOIprefix\doi{10.1111/j.1365-2966.2011.18794.x},
  \href{http://arxiv.org/abs/1101.1961}{\tt arXiv:1101.1961}.
%Type = Article
\bibitem[{{Ma} et~al.(2018){Ma}, {Angryk}, {Riley} and {Filali
  Boubrahimi}}]{2018Ma}
\bibinfo{author}{{Ma}, R.}, \bibinfo{author}{{Angryk}, R.A.},
  \bibinfo{author}{{Riley}, P.}, \bibinfo{author}{{Filali Boubrahimi}, S.},
  \bibinfo{year}{2018}.
\newblock \bibinfo{title}{{Coronal Mass Ejection Data Clustering and
  Visualization of Decision Trees}}.
\newblock \bibinfo{journal}{Astrophysics Journals} \bibinfo{volume}{236},
  \bibinfo{pages}{14}.
\newblock \DOIprefix\doi{10.3847/1538-4365/aab76f}.
%Type = Article
\bibitem[{{Marchi}(2007)}]{2007ApJ...666..475M}
\bibinfo{author}{{Marchi}, S.}, \bibinfo{year}{2007}.
\newblock \bibinfo{title}{{Extrasolar Planet Taxonomy: A New Statistical
  Approach}}.
\newblock \bibinfo{journal}{Astrophysics Journal} \bibinfo{volume}{666},
  \bibinfo{pages}{475--485}.
\newblock \DOIprefix\doi{10.1086/519760},
  \href{http://arxiv.org/abs/0705.0910}{\tt arXiv:0705.0910}.
%Type = Article
\bibitem[{{Marchi} et~al.(2009){Marchi}, {Ortolani}, {Nagasawa} and
  {Ida}}]{2009Marchi}
\bibinfo{author}{{Marchi}, S.}, \bibinfo{author}{{Ortolani}, S.},
  \bibinfo{author}{{Nagasawa}, M.}, \bibinfo{author}{{Ida}, S.},
  \bibinfo{year}{2009}.
\newblock \bibinfo{title}{{On the various origins of close-in extrasolar
  planets}}.
\newblock \bibinfo{journal}{Monthly Notices of the Royal Astronomy Society}
  \bibinfo{volume}{394}, \bibinfo{pages}{L93--L96}.
\newblock \DOIprefix\doi{10.1111/j.1745-3933.2009.00619.x},
  \href{http://arxiv.org/abs/0901.1547}{\tt arXiv:0901.1547}.
%Type = Article
\bibitem[{{Materne}(1978)}]{1978A&A....63..401M}
\bibinfo{author}{{Materne}, J.}, \bibinfo{year}{1978}.
\newblock \bibinfo{title}{{The structure of nearby clusters of galaxies.
  Hierarchical clustering and an application to the Leo region.}}
\newblock \bibinfo{journal}{Astronomy and Astrophysics} \bibinfo{volume}{63},
  \bibinfo{pages}{401--409}.
%Type = Article
\bibitem[{{Milani} et~al.(2014){Milani}, {Cellino}, {Kne{\v{z}}evi{\'c}},
  {Novakovi{\'c}}, {Spoto} and {Paolicchi}}]{2014Milani}
\bibinfo{author}{{Milani}, A.}, \bibinfo{author}{{Cellino}, A.},
  \bibinfo{author}{{Kne{\v{z}}evi{\'c}}, Z.}, \bibinfo{author}{{Novakovi{\'c}},
  B.}, \bibinfo{author}{{Spoto}, F.}, \bibinfo{author}{{Paolicchi}, P.},
  \bibinfo{year}{2014}.
\newblock \bibinfo{title}{{Asteroid families classification: Exploiting very
  large datasets}}.
\newblock \bibinfo{journal}{Icarus} \bibinfo{volume}{239},
  \bibinfo{pages}{46--73}.
\newblock \DOIprefix\doi{10.1016/j.icarus.2014.05.039},
  \href{http://arxiv.org/abs/1312.7702}{\tt arXiv:1312.7702}.
%Type = Article
\bibitem[{{Milani} et~al.(2019){Milani}, {Kne{\v{z}}evi{\'c}}, {Spoto} and
  {Paolicchi}}]{2019Milani}
\bibinfo{author}{{Milani}, A.}, \bibinfo{author}{{Kne{\v{z}}evi{\'c}}, Z.},
  \bibinfo{author}{{Spoto}, F.}, \bibinfo{author}{{Paolicchi}, P.},
  \bibinfo{year}{2019}.
\newblock \bibinfo{title}{{Asteroid cratering families: recognition and
  collisional interpretation}}.
\newblock \bibinfo{journal}{Astronomy and Astrophysics} \bibinfo{volume}{622},
  \bibinfo{pages}{A47}.
\newblock \DOIprefix\doi{10.1051/0004-6361/201834056},
  \href{http://arxiv.org/abs/1812.07535}{\tt arXiv:1812.07535}.
%Type = Article
\bibitem[{{Peth} et~al.(2016){Peth}, {Lotz}, {Freeman}, {McPartland},
  {Mortazavi}, {Snyder}, {Barro}, {Grogin}, {Guo}, {Hemmati}, {Kartaltepe},
  {Kocevski}, {Koekemoer}, {McIntosh}, {Nayyeri}, {Papovich}, {Primack} and
  {Simons}}]{2016Peth}
\bibinfo{author}{{Peth}, M.A.}, \bibinfo{author}{{Lotz}, J.M.},
  \bibinfo{author}{{Freeman}, P.E.}, \bibinfo{author}{{McPartland}, C.},
  \bibinfo{author}{{Mortazavi}, S.A.}, \bibinfo{author}{{Snyder}, G.F.},
  \bibinfo{author}{{Barro}, G.}, \bibinfo{author}{{Grogin}, N.A.},
  \bibinfo{author}{{Guo}, Y.}, \bibinfo{author}{{Hemmati}, S.},
  \bibinfo{author}{{Kartaltepe}, J.S.}, \bibinfo{author}{{Kocevski}, D.D.},
  \bibinfo{author}{{Koekemoer}, A.M.}, \bibinfo{author}{{McIntosh}, D.H.},
  \bibinfo{author}{{Nayyeri}, H.}, \bibinfo{author}{{Papovich}, C.},
  \bibinfo{author}{{Primack}, J.R.}, \bibinfo{author}{{Simons}, R.C.},
  \bibinfo{year}{2016}.
\newblock \bibinfo{title}{{Beyond spheroids and discs: classifications of
  CANDELS galaxy structure at 1.4 < z < 2 via principal component analysis}}.
\newblock \bibinfo{journal}{Monthly Notices of the Royal Astronomy Society}
  \bibinfo{volume}{458}, \bibinfo{pages}{963--987}.
\newblock \DOIprefix\doi{10.1093/mnras/stw252},
  \href{http://arxiv.org/abs/1504.01751}{\tt arXiv:1504.01751}.
%Type = Article
\bibitem[{{Rice} et~al.(2016){Rice}, {Goodman}, {Bergin}, {Beaumont} and
  {Dame}}]{2016ApJ...822...52R}
\bibinfo{author}{{Rice}, T.S.}, \bibinfo{author}{{Goodman}, A.A.},
  \bibinfo{author}{{Bergin}, E.A.}, \bibinfo{author}{{Beaumont}, C.},
  \bibinfo{author}{{Dame}, T.M.}, \bibinfo{year}{2016}.
\newblock \bibinfo{title}{{A Uniform Catalog of Molecular Clouds in the Milky
  Way}}.
\newblock \bibinfo{journal}{Astrophysics Journal} \bibinfo{volume}{822},
  \bibinfo{pages}{52}.
\newblock \DOIprefix\doi{10.3847/0004-637X/822/1/52},
  \href{http://arxiv.org/abs/1602.02791}{\tt arXiv:1602.02791}.
%Type = Inproceedings
\bibitem[{{Rood}(1983)}]{1983Rood}
\bibinfo{author}{{Rood}, H.J.}, \bibinfo{year}{1983}.
\newblock \bibinfo{title}{{Dendogram Cosmography - the Stars Within 25
  Parsecs}}, in: \bibinfo{editor}{{Philip}, A.G.D.}, \bibinfo{editor}{{Upgren},
  A.R.} (Eds.), \bibinfo{booktitle}{IAU Colloq. 76: Nearby Stars and the
  Stellar Luminosity Function}, p. \bibinfo{pages}{411}.
%Type = Article
\bibitem[{{Rosolowsky} et~al.(2008){Rosolowsky}, {Pineda}, {Kauffmann} and
  {Goodman}}]{2008ApJ...679.1338R}
\bibinfo{author}{{Rosolowsky}, E.W.}, \bibinfo{author}{{Pineda}, J.E.},
  \bibinfo{author}{{Kauffmann}, J.}, \bibinfo{author}{{Goodman}, A.A.},
  \bibinfo{year}{2008}.
\newblock \bibinfo{title}{{Structural Analysis of Molecular Clouds:
  Dendrograms}}.
\newblock \bibinfo{journal}{Astrophysics Journal} \bibinfo{volume}{679},
  \bibinfo{pages}{1338--1351}.
\newblock \DOIprefix\doi{10.1086/587685},
  \href{http://arxiv.org/abs/0802.2944}{\tt arXiv:0802.2944}.
%Type = Article
\bibitem[{{Sampedro} and {Alfaro}(2016)}]{2016MNRAS.457.3949S}
\bibinfo{author}{{Sampedro}, L.}, \bibinfo{author}{{Alfaro}, E.J.},
  \bibinfo{year}{2016}.
\newblock \bibinfo{title}{{Stellar open clusters' membership probabilities: an
  N-dimensional geometrical approach}}.
\newblock \bibinfo{journal}{Monthly Notices of the Royal Astronomy Society}
  \bibinfo{volume}{457}, \bibinfo{pages}{3949--3962}.
\newblock \DOIprefix\doi{10.1093/mnras/stw243},
  \href{http://arxiv.org/abs/1602.01025}{\tt arXiv:1602.01025}.
%Type = Article
\bibitem[{{Sanders}(1971)}]{1971A&A....14..226S}
\bibinfo{author}{{Sanders}, W.L.}, \bibinfo{year}{1971}.
\newblock \bibinfo{title}{{An improved method for computing membership
  probabilities in open clusters.}}
\newblock \bibinfo{journal}{Astronomy and Astrophysics} \bibinfo{volume}{14},
  \bibinfo{pages}{226--232}.
%Type = Article
\bibitem[{{Santiago-Bautista} et~al.(2020){Santiago-Bautista}, {Caretta},
  {Bravo-Alfaro}, {Pointecouteau} and {Andernach}}]{2020A&A...637A..31S}
\bibinfo{author}{{Santiago-Bautista}, I.}, \bibinfo{author}{{Caretta}, C.A.},
  \bibinfo{author}{{Bravo-Alfaro}, H.}, \bibinfo{author}{{Pointecouteau}, E.},
  \bibinfo{author}{{Andernach}, H.}, \bibinfo{year}{2020}.
\newblock \bibinfo{title}{{Identification of filamentary structures in the
  environment of superclusters of galaxies in the Local Universe}}.
\newblock \bibinfo{journal}{Astronomy and Astrophysics} \bibinfo{volume}{637},
  \bibinfo{pages}{A31}.
\newblock \DOIprefix\doi{10.1051/0004-6361/201936397},
  \href{http://arxiv.org/abs/2002.03446}{\tt arXiv:2002.03446}.
%Type = Article
\bibitem[{{Sarro} et~al.(2014){Sarro}, {Bouy}, {Berihuete}, {Bertin}, {Moraux},
  {Bouvier}, {Cuillandre}, {Barrado} and {Solano}}]{2014A&A...563A..45S}
\bibinfo{author}{{Sarro}, L.M.}, \bibinfo{author}{{Bouy}, H.},
  \bibinfo{author}{{Berihuete}, A.}, \bibinfo{author}{{Bertin}, E.},
  \bibinfo{author}{{Moraux}, E.}, \bibinfo{author}{{Bouvier}, J.},
  \bibinfo{author}{{Cuillandre}, J.C.}, \bibinfo{author}{{Barrado}, D.},
  \bibinfo{author}{{Solano}, E.}, \bibinfo{year}{2014}.
\newblock \bibinfo{title}{{Cluster membership probabilities from proper motions
  and multi-wavelength photometric catalogues. I. Method and application to the
  Pleiades cluster}}.
\newblock \bibinfo{journal}{Astronomy and Astrophysics} \bibinfo{volume}{563},
  \bibinfo{pages}{A45}.
\newblock \DOIprefix\doi{10.1051/0004-6361/201322413},
  \href{http://arxiv.org/abs/1401.7427}{\tt arXiv:1401.7427}.
%Type = Article
\bibitem[{{Schmeja}(2011)}]{2011Schmeja}
\bibinfo{author}{{Schmeja}, S.}, \bibinfo{year}{2011}.
\newblock \bibinfo{title}{{Identifying star clusters in a field: A comparison
  of different algorithms}}.
\newblock \bibinfo{journal}{Astronomische Nachrichten} \bibinfo{volume}{332},
  \bibinfo{pages}{172}.
\newblock \DOIprefix\doi{10.1002/asna.201011484},
  \href{http://arxiv.org/abs/1011.5533}{\tt arXiv:1011.5533}.
%Type = Article
\bibitem[{{Serna} and {Gerbal}(1996)}]{1996A&A...309...65S}
\bibinfo{author}{{Serna}, A.}, \bibinfo{author}{{Gerbal}, D.},
  \bibinfo{year}{1996}.
\newblock \bibinfo{title}{{Dynamical search for substructures in galaxy
  clusters. A hierarchical clustering method.}}
\newblock \bibinfo{journal}{Astronomy and Astrophysics} \bibinfo{volume}{309},
  \bibinfo{pages}{65--74}.
\newblock \href{http://arxiv.org/abs/astro-ph/9509080}{\tt
  arXiv:astro-ph/9509080}.
%Type = Article
\bibitem[{{Serra} and {Diaferio}(2013)}]{2013ApJ...768..116S}
\bibinfo{author}{{Serra}, A.L.}, \bibinfo{author}{{Diaferio}, A.},
  \bibinfo{year}{2013}.
\newblock \bibinfo{title}{{Identification of Members in the Central and Outer
  Regions of Galaxy Clusters}}.
\newblock \bibinfo{journal}{Astrophysics Journal} \bibinfo{volume}{768},
  \bibinfo{pages}{116}.
\newblock \DOIprefix\doi{10.1088/0004-637X/768/2/116},
  \href{http://arxiv.org/abs/1211.3669}{\tt arXiv:1211.3669}.
%Type = Article
\bibitem[{Sibson(1973)}]{Sibson1973SLINKAO}
\bibinfo{author}{Sibson, R.}, \bibinfo{year}{1973}.
\newblock \bibinfo{title}{Slink: An optimally efficient algorithm for the
  single-link cluster method}.
\newblock \bibinfo{journal}{Computer Journal} \bibinfo{volume}{16},
  \bibinfo{pages}{30--34}.
%Type = Article
\bibitem[{{Smullen} et~al.(2020){Smullen}, {Kratter}, {Offner}, {Lee} and
  {Chen}}]{2020MNRAS.497.4517S}
\bibinfo{author}{{Smullen}, R.A.}, \bibinfo{author}{{Kratter}, K.M.},
  \bibinfo{author}{{Offner}, S.S.R.}, \bibinfo{author}{{Lee}, A.T.},
  \bibinfo{author}{{Chen}, H.H.H.}, \bibinfo{year}{2020}.
\newblock \bibinfo{title}{{The highly variable time evolution of star-forming
  cores identified with dendrograms}}.
\newblock \bibinfo{journal}{Monthly Notices of the Royal Astronomy Society}
  \bibinfo{volume}{497}, \bibinfo{pages}{4517--4534}.
\newblock \DOIprefix\doi{10.1093/mnras/staa2253},
  \href{http://arxiv.org/abs/2004.01263}{\tt arXiv:2004.01263}.
%Type = Inproceedings
\bibitem[{S{\o}rensen(1948)}]{Sorensen1948}
\bibinfo{author}{S{\o}rensen, T.}, \bibinfo{year}{1948}.
\newblock \bibinfo{title}{A method of establishing group of equal amplitude in
  plant sociobiology based on similarity of species content and its application
  to analyses of the vegetation on danish commons}, in:
  \bibinfo{booktitle}{Biologiske Skrifter}.
%Type = Article
\bibitem[{Tully(1980)}]{Tully1980}
\bibinfo{author}{Tully, R.B.}, \bibinfo{year}{1980}.
\newblock \bibinfo{title}{Nearby groups of galaxies. {I} - {The} {NGC} 1023
  group}.
\newblock \bibinfo{journal}{The Astrophysical Journal} \bibinfo{volume}{237},
  \bibinfo{pages}{390--403}.
\newblock \URLprefix \url{http://adsabs.harvard.edu/abs/1980ApJ...237..390T},
  \DOIprefix\doi{10.1086/157881}.
%Type = Article
\bibitem[{Tully(1987)}]{Tully1987}
\bibinfo{author}{Tully, R.B.}, \bibinfo{year}{1987}.
\newblock \bibinfo{title}{Nearby groups of galaxies. {II} - an all-sky survey
  within 3000 kilometers per second}.
\newblock \bibinfo{journal}{The Astrophysical Journal} \bibinfo{volume}{321},
  \bibinfo{pages}{280--304}.
\newblock \URLprefix \url{http://adsabs.harvard.edu/abs/1987ApJ...321..280T},
  \DOIprefix\doi{10.1086/165629}.
%Type = Article
\bibitem[{{Vasilevskis} et~al.(1958){Vasilevskis}, {Klemola} and
  {Preston}}]{1958AJ.....63..387V}
\bibinfo{author}{{Vasilevskis}, S.}, \bibinfo{author}{{Klemola}, A.},
  \bibinfo{author}{{Preston}, G.}, \bibinfo{year}{1958}.
\newblock \bibinfo{title}{{Relative proper motions of stars in the region of
  the open cluster NGC 6633.}}
\newblock \bibinfo{journal}{Astronomical Journal} \bibinfo{volume}{63},
  \bibinfo{pages}{387--395}.
\newblock \DOIprefix\doi{10.1086/107787}.
%Type = Article
\bibitem[{{WANG} and {YU}(2021)}]{wang2021}
\bibinfo{author}{{WANG}, L.}, \bibinfo{author}{{YU}, H.}, \bibinfo{year}{2021}.
\newblock \bibinfo{title}{On merging galaxy cluster macs j0358.8-2955}.
\newblock \bibinfo{journal}{Beijing Normal University(Natural Science)}
  \bibinfo{volume}{57}, \bibinfo{pages}{186--193}.
%Type = Article
\bibitem[{Wilkinson and Friendly(2009)}]{2009:HCHM}
\bibinfo{author}{Wilkinson, L.}, \bibinfo{author}{Friendly, M.},
  \bibinfo{year}{2009}.
\newblock \bibinfo{title}{The history of the cluster heat map}.
\newblock \bibinfo{journal}{The American Statistician} \bibinfo{volume}{63},
  \bibinfo{pages}{179--184}.
%Type = Article
\bibitem[{Yu et~al.(2016)Yu, Diaferio, Agulli, Aguerri and Tozzi}]{Yu2016}
\bibinfo{author}{Yu, H.}, \bibinfo{author}{Diaferio, A.},
  \bibinfo{author}{Agulli, I.}, \bibinfo{author}{Aguerri, J.A.L.},
  \bibinfo{author}{Tozzi, P.}, \bibinfo{year}{2016}.
\newblock \bibinfo{title}{The unrelaxed dynamical structure of the galaxy
  cluster abell 85}.
\newblock \bibinfo{journal}{The Astrophysical Journal} \bibinfo{volume}{831}.
\newblock \URLprefix \url{http://adsabs.harvard.edu/abs/2016ApJ...831..156Y}.
%Type = Article
\bibitem[{{Yu} et~al.(2018){Yu}, {Diaferio}, {Serra} and {Baldi}}]{Yu2018b}
\bibinfo{author}{{Yu}, H.}, \bibinfo{author}{{Diaferio}, A.},
  \bibinfo{author}{{Serra}, A.L.}, \bibinfo{author}{{Baldi}, M.},
  \bibinfo{year}{2018}.
\newblock \bibinfo{title}{{Blooming Trees: Substructures and Surrounding Groups
  of Galaxy Clusters}}.
\newblock \bibinfo{journal}{Astrophysics Journal} \bibinfo{volume}{860},
  \bibinfo{pages}{118}.
\newblock \DOIprefix\doi{10.3847/1538-4357/aac263},
  \href{http://arxiv.org/abs/1805.12306}{\tt arXiv:1805.12306}.
%Type = Article
\bibitem[{{Yu} et~al.(2015){Yu}, {Serra}, {Diaferio} and
  {Baldi}}]{2015ApJ...810...37Y}
\bibinfo{author}{{Yu}, H.}, \bibinfo{author}{{Serra}, A.L.},
  \bibinfo{author}{{Diaferio}, A.}, \bibinfo{author}{{Baldi}, M.},
  \bibinfo{year}{2015}.
\newblock \bibinfo{title}{{Identification of Galaxy Cluster Substructures with
  the Caustic Method}}.
\newblock \bibinfo{journal}{Astrophysics Journal} \bibinfo{volume}{810},
  \bibinfo{pages}{37}.
\newblock \DOIprefix\doi{10.1088/0004-637X/810/1/37},
  \href{http://arxiv.org/abs/1503.08823}{\tt arXiv:1503.08823}.
%Type = Article
\bibitem[{{Yu} et~al.(2020){Yu}, {Shao}, {Diaferio} and {Li}}]{2020Yu}
\bibinfo{author}{{Yu}, H.}, \bibinfo{author}{{Shao}, Z.},
  \bibinfo{author}{{Diaferio}, A.}, \bibinfo{author}{{Li}, L.},
  \bibinfo{year}{2020}.
\newblock \bibinfo{title}{{Unveiling the Hierarchical Structure of Open Star
  Clusters: The Perseus Double Cluster}}.
\newblock \bibinfo{journal}{Astrophysics Journal} \bibinfo{volume}{899},
  \bibinfo{pages}{144}.
\newblock \DOIprefix\doi{10.3847/1538-4357/aba8f3},
  \href{http://arxiv.org/abs/2007.11850}{\tt arXiv:2007.11850}.
%Type = Article
\bibitem[{{Zappal{\`a}} et~al.(1995){Zappal{\`a}}, {Bendjoya}, {Cellino},
  {Farinella} and {Froeschl{\'e}}}]{1995Icar..116..291Z}
\bibinfo{author}{{Zappal{\`a}}, V.}, \bibinfo{author}{{Bendjoya}, P.},
  \bibinfo{author}{{Cellino}, A.}, \bibinfo{author}{{Farinella}, P.},
  \bibinfo{author}{{Froeschl{\'e}}, C.}, \bibinfo{year}{1995}.
\newblock \bibinfo{title}{{Asteroid families: Search of a 12,487-asteroid
  sample using two different clustering techniques.}}
\newblock \bibinfo{journal}{Icarus} \bibinfo{volume}{116},
  \bibinfo{pages}{291--314}.
\newblock \DOIprefix\doi{10.1006/icar.1995.1127}.
%Type = Article
\bibitem[{{Zappala} et~al.(1990){Zappala}, {Cellino}, {Farinella} and
  {Knezevic}}]{1990Zappala}
\bibinfo{author}{{Zappala}, V.}, \bibinfo{author}{{Cellino}, A.},
  \bibinfo{author}{{Farinella}, P.}, \bibinfo{author}{{Knezevic}, Z.},
  \bibinfo{year}{1990}.
\newblock \bibinfo{title}{{Asteroid Families. I. Identification by Hierarchical
  Clustering and Reliability Assessment}}.
\newblock \bibinfo{journal}{Astronomical Journal} \bibinfo{volume}{100},
  \bibinfo{pages}{2030}.
\newblock \DOIprefix\doi{10.1086/115658}.
%Type = Article
\bibitem[{{Zappala} et~al.(1994){Zappala}, {Cellino}, {Farinella} and
  {Milani}}]{1994Zappala}
\bibinfo{author}{{Zappala}, V.}, \bibinfo{author}{{Cellino}, A.},
  \bibinfo{author}{{Farinella}, P.}, \bibinfo{author}{{Milani}, A.},
  \bibinfo{year}{1994}.
\newblock \bibinfo{title}{{Asteroid Famalies. II. Extension to Unnumbered
  Multiopposition Asteroids}}.
\newblock \bibinfo{journal}{Astronomical Journal} \bibinfo{volume}{107},
  \bibinfo{pages}{772}.
\newblock \DOIprefix\doi{10.1086/116897}.
%Type = Article
\bibitem[{{Zhao} and {He}(1990)}]{1990A&A...237...54Z}
\bibinfo{author}{{Zhao}, J.L.}, \bibinfo{author}{{He}, Y.P.},
  \bibinfo{year}{1990}.
\newblock \bibinfo{title}{{An improved method for membership determination of
  stellar clusters with proper motions with different accuracies.}}
\newblock \bibinfo{journal}{Astronomy and Astrophysics} \bibinfo{volume}{237},
  \bibinfo{pages}{54}.

\end{thebibliography}

\end{document}